\documentclass[%
 reprint,
superscriptaddress,
nofootinbib,
 amsmath,amssymb,
 aps,
prd,
floatfix,
longbibliography
]{revtex4-1}

\usepackage{graphicx}
\usepackage{dcolumn}
\usepackage{bm}
\usepackage{hyperref}
\usepackage[dvipsnames]{xcolor}
\usepackage[mathlines]{lineno}
\usepackage{tabularx}
\usepackage{subfigure}
\usepackage{multirow}
\usepackage[normalem]{ulem}
\usepackage[export]{adjustbox}
\usepackage{url}
\usepackage{academicons}

\hypersetup{
    pdfnewwindow=true,    
    colorlinks=true,      
    linkcolor=blue,       
    citecolor=blue,       
    filecolor=blue,       
    urlcolor=blue         
}

\newcolumntype{Y}{>{\centering\arraybackslash}X}

\usepackage{natbib}

\begin{document}
\title{Searching~for~Particle~Dark~Matter~with~eROSITA~Early~Data}

\author{Chingam Fong}
\email{fongchingam@link.cuhk.edu.hk}
\thanks{\scriptsize \!\! \href{http://orcid.org/0009-0002-3492-9849}{orcid.org/0009-0002-3492-9849}}
\affiliation{
Department of Physics, The Chinese University of Hong Kong, Shatin, Hong Kong, China}

\author{Kenny C. Y. Ng}
\email{kcyng@phy.cuhk.edu.hk}
\thanks{\scriptsize \!\! \href{http://orcid.org/0000-0001-8016-2170}{orcid.org/0000-0001-8016-2170}}
\affiliation{
Department of Physics, The Chinese University of Hong Kong, Shatin, Hong Kong, China}

\author{Qishan Liu}
\email{qisliu@link.cuhk.edu.hk}
\thanks{\scriptsize \!\! \href{http://orcid.org/0000-0003-1437-6829}{orcid.org/0000-0003-1437-6829}}
\affiliation{
Department of Physics, The Chinese University of Hong Kong, Shatin, Hong Kong, China}
\affiliation{Institute of High Energy Physics, Chinese Academy of Sciences, Beijing 1000049, China}
\affiliation{China Center of Advanced Science and Technology, Beijing 100190, China}

\date{September 6, 2025}


\begin{abstract}

Many well motivated dark matter (DM) particle candidates can decay into detectable X-ray photons. We analyze the Final Equatorial Depth Survey (eFEDS) data from eROSITA early data release to search for unidentified X-ray lines that could indicate DM signals. Having discovered no anomalous signal, we set limits on DM decay rate in mass range between 1.8-18 keV, and constrain the parameter space of two DM particles: sterile neutrino and Axion-like particle. Finally we also study the projected sensitivity of eROSITA full sky search, showing that eROSITA all-sky survey is expected to set the most stringent limits in the soft X-ray band.

\end{abstract}


\maketitle

\section{\label{sec:intro}Introduction}

\par The evidences of Dark Matter (DM, symbolized with subscript $\chi$ in equations) exist on all scales~\cite{ParticleDataGroup:2022pth}. From galaxy rotation curve, to the power spectrum of large scale structure and anisotropy of CMB, all hint that $\sim26\%$ of the energy content of the universe is non-baryon and non-luminous~\cite{Bertone:2004pz, PerezdelosHeros:2020qyt, Strigari:2013iaa, 10.1088/978-1-6817-4118-5, Buckley:2017ijx, Planck:2018vyg, Wechsler_2018, Salucci:2018hqu}. Identifying the nature of DM remains one of the most important problems in modern physics.

\par One popular way for testing the particle nature of DM is Indirect Detection. The principle is simple: if DM particles can decay/annihilate into Standard Model particles, an observable signature will show up in astrophysical surveys~\cite{Gaskins_2016}. The strength of the signal is proportional to DM density (or density squared for annihilation) in the field of view and the decay/annihilation rate. Conversely, if said signal is absent in astrophysical observations, the survey can be used to constrain the DM theory. 

\par There has been a lot of effort on using X-ray to test for light DM particles.~\cite{Horiuchi:2013noa, Urban:2014yda, Malyshev:2014xqa, Anderson:2014tza, Tamura:2014mta, Jeltema:2014qfa, Jeltema:2015mee, Aharonian:2016gzq, Gewering_Peine_2017, Silich:2021sra, Sicilian:2020glg, Sicilian:2022wvm, Dessert:2018qih, Foster_2021, Roach_2023, Watson:2006qb, Boyarsky:2006fg, Boyarsky:2007ay, Boyarsky:2007ge, Yuksel:2007xh, Loewenstein:2008yi, RiemerSorensen:2009jp, Horiuchi:2013noa, Urban:2014yda, Tamura:2014mta, Figueroa-Feliciano:2015gwa, Ng:2015gfa, Iakubovskyi:2015dna, Aharonian:2016gzq, Sekiya:2015jsa, Dessert:2018qih, Roach_2023, Hofmann:2016urz, Sicilian:2020glg, Foster_2021, Silich:2021sra, Bhargava:2020fxr}, as various particle DM scenarios have predicted signals in the X-ray range. E.g., DM annihilation or decay into primary or secondary electrons, which can produce MeV scale emission through inverse Compton scattering~\cite{Cirelli:2023tnx}; sterile neutrinos or Axions could decay into monoenergetic photons with energy being half of their mass~\cite{Bertone:2010zza}; composite dark atoms excitation and de-excitation to produce MeV gamma ray excess of the galactic bulge~\cite{Khlopov:2008ki, Khlopov:2010pq, Cudell:2014jba, Cudell:2014wca}. To test these models, there remains a need for more analysis employing new observations, instrument, and techniques. 

\par X-ray searches of DM had gained significant attention in the past decade, when a stacked X-ray analysis of XMM-Newton studying 73 galactic clusters indicated there’s an unexplained line around $3.5\rm\,keV$, which could correspond to the decay signature of a DM particle with mass at $7\rm\,keV$~\cite{Bulbul:2014sua}. $3.5\rm\,keV$ line was also claimed in 3 later studies: XMM-Newton observation of Andromeda and Perseus cluster~\cite{Boyarsky:2014jta}, XMM-Newton observation of Milky Way Center~\cite{Boyarsky:2014ska}, and Chandra deep field observation~\cite{Cappelluti:2017ywp}. However, other studies could not find $3.5\rm\,keV$ line or any other indication of DM decay~\cite{Horiuchi:2013noa, Urban:2014yda, Malyshev:2014xqa, Anderson:2014tza, Tamura:2014mta, Jeltema:2014qfa, Jeltema:2015mee, Aharonian:2016gzq, Gewering_Peine_2017, Silich:2021sra, Sicilian:2020glg, Sicilian:2022wvm, Dessert:2018qih, Foster_2021, Roach_2023}. Recently, a study~\cite{Dessert:2023fen} attempted to reproduce the 3 original analyses where $3.5\,\rm keV$ was discovered, and could not NFWfound robust evidence to support its existence. Nevertheless, X-ray line searches remain a important way of searching for DM signatures.

\par Extended ROentgen Survey with an Imaging Telescope Array (eROSITA) is one of two X-ray telescopes onboard observatory satellite Spectrum-Roentgen-Gamma (SRG). Started in December 2019, the mission of eROSITA is to provide the deepest view of the whole $4\pi$ X-ray sky. Once all 8 planned scans are completed, eROSITA all-sky survey will provide the first true whole sky imaging in hard X-ray band ($\rm2.3$-$8\rm\,keV$). While in soft X-ray band ($\rm0.2$-$2.3\rm\,keV$) it will be 25 times more sensitive than its predecessor, the ROSAT all-sky survey~\cite{eROSITA:2020emt}. Projections using pre-launch estimations parameters show that, with its excellent angular and energy resolution, large field of view, and long exposure time, eROSITA survey has the potential to set new limits in X-ray DM search~\cite{Dekker:2021bos, Barinov:2022kfp}.

\par In this paper, we perform indirect DM search on the largest continuous observation from publicly available eROSITA data. We search for DM decay signal in eROSITA Equatorial Field Depth Survey (eFEDS) data. In the event of null detection, we set upper limit on the decay rate of DM. We also convert the decay rate limit to parameter space constraints for sterile neutrino mixing angle and Axion-Like Particles couplings in mass range $2$-$18\rm\,keV$. Finally, we extrapolate the result that would be obtained from eROSITA using 4 years of full sky data. This study also serves as a preparation for DM analysis on eROSITA full data release. 

\par This paper is structured as follows, in section~\ref{sec:analysis} we describe the instrument of eROSITA, the data set, and the background model we employed. In section~\ref{sec:dm_analysis} we present our DM constraint derived from eFEDS, convert this limit to constrain both sterile neutrino and Axion Like Particle parameter space, and project future results from eROSITA. We conclude the paper in section~\ref{sec:conclusions} by discussing the quirks of the result and the possible future developments in eROSITA DM study. %

\section{\label{sec:analysis}\MakeLowercase{e}ROSITA Instrument and Data}

\subsection{\label{sec:instrument}eROSITA Instrument}

\par eROSITA consists of 7 independent telescope modules (TM1-TM7), arranged in a honey comb like geometry~\cite{2018SPIE10699E..1ZE}. A TM contains a mirror assembly and a camera module. The mirror assembly has 54 paraboloid and hyperboloid mirror shells in a Wolter-I geometry. There’s a camera at the focus of the mirror. It has a circular field of view with diameter at 1.03 degrees, which makes the geometric area of the detector 0.833 deg$^2$. The grasp of eROSITA, defined as the energy dependent multiplied product of effective area and field of view, is competitive in 3.5-10 keV range, while surpassing other instruments in 0.3-3.5 keV~\cite{eROSITA:2020emt}. 

\par Each TM has a separate light blocking Aluminum filter for protection against unwanted optical light. A 200 nm Al layer is deposited directly on camera for TM1, 2, 3, 4, 6, while TM5 and TM7 has an external 100 nm Al layer on the filter wheel, since they were planned to use for low energy spectroscopy. After mission started, it was noticed some optical light bypassed TM5 and 7’s filter and contaminated the spectrum (light leak, see section 9.2 of~\cite{eROSITA:2020emt}). 

\par Since the light leak was caused by optical filter failing, the contaminated energy range is limited to below $\sim$ 1 keV. It was noticed the light leak heavily depends on SRG's angle relative to the Sun. Moreover, the affected region is limited to the bottom part of the detector. Therefore, one could investigate a thorough treatment of the light leak issue, either by setting bad time interval or setting bad pixel region and remove them in data processing. 

\par For this work, we ignore the energy channels below 0.9 keV in our spectral fitting procedure and in our DM analysis. This solution is straightforward but effective. In doing so, we make use of data taken by all 7 modules, while mitigate light leak effect such that our blank sky spectrum is still valid.

\subsection{\label{sec:eFEDS}eROSITA Early Data eFEDS and Software}

\begin{table}
\centering
\begin{tabularx}{0.45\textwidth}{YYYYY}
\hline
\hline
eROSITA obsID & Central RA & Central DEC & Effective Exposure\footnote{during this time all 7 TMs were working} \\
 & (deg) & (deg) & (s) \\
 \hline
300007-1 & 129.5500 & 1.5000  & 89642.0 \\
300008-1 & 133.8600 & 1.5000  & 89642.0 \\
300009-1 & 138.1400 & 1.5000  & 89642.0 \\
300010-1 & 142.4500 & 1.5000  & 89642.0 \\
 \hline
 \hline
\end{tabularx}%
\caption{eROSITA observations that make up four eFEDS sub-fields, which are used in our analysis.}
 \label{obs_table}
\end{table}

\begin{figure*}[t]
    \centering
    \includegraphics[width=\textwidth]{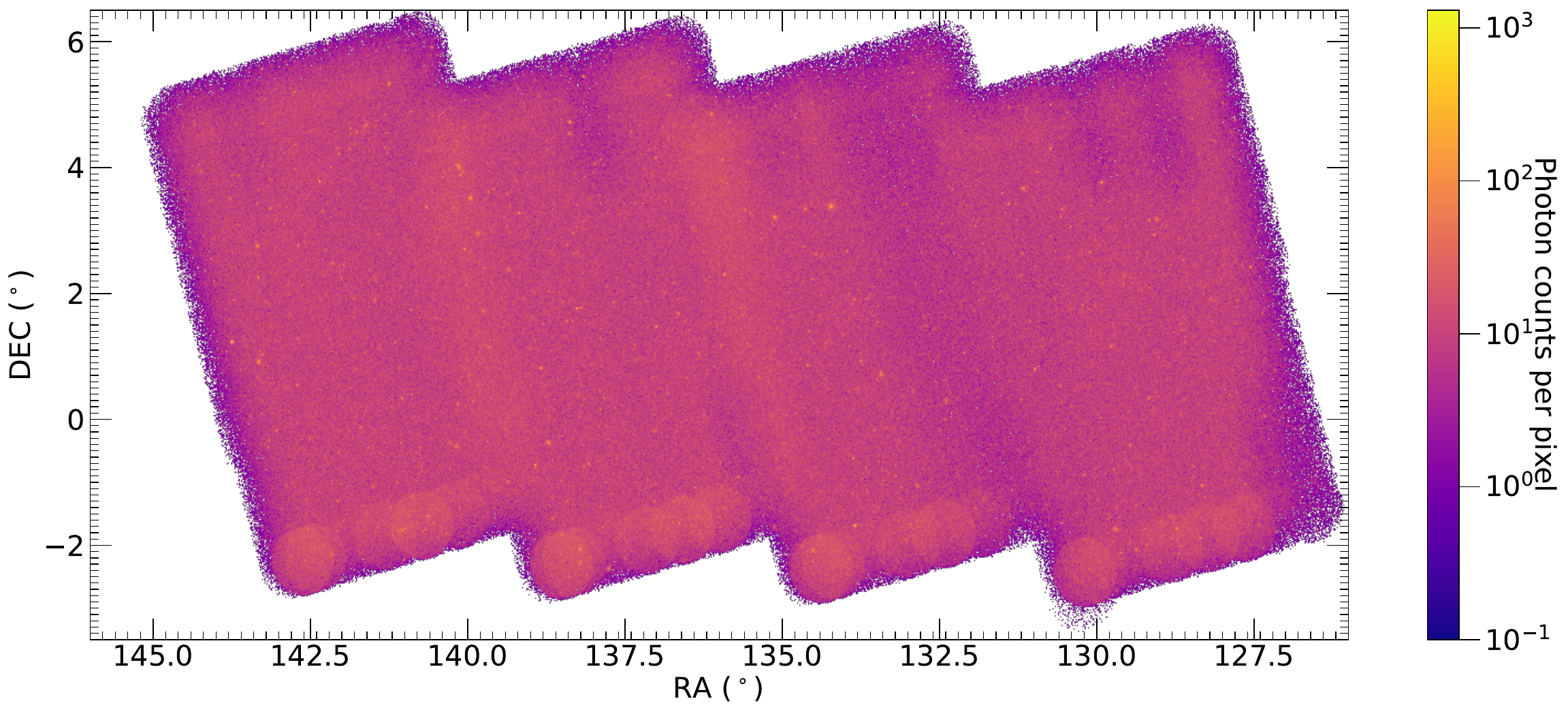}
    \caption{X-ray sky image of the observation eFEDS, generated with \texttt{evtool}. Each pixel on this picture has a width of $64''$. }
    \label{fig:sky}
\end{figure*}

\par 
Before all-sky survey began, eROSITA completed a number of Calibration and Performance Verification (Cal-PV) observations in late 2019. In June 2021, the portion of Cal-PV observations proposed by German eROSITA team were released to the public as eROSITA Early Data Release (EDR) \footnote{data download available at \url{https://erosita.mpe.mpg.de/edr/eROSITAObservations/}}.

\par Figure~\ref{fig:sky} shows a sky image of the eROSITA Equatorial Field Depth Survey (eFEDS) observation, the data set we have chosen to use to search for potential DM signature. eFEDS observed a region centered at ${\rm RA}=136^\circ, {\rm DEC}=1.5^\circ$ for a total of $\sim360$ ks with all 7 telescope modules. The observation started at the beginning of November 2019. Of all the released eROSITA early data, it has the longest observation time and largest continuous observation area, making it the most suitable available eROSITA data for indirect DM search. See table~\ref{obs_table} for details on 4 eFEDS observations.  

\par Alongside observed data, EDR also made available an analysis software to extract and manipulate eROSITA data, named eROSITA Science Analysis Software System (eSASS) \footnote{\url{https://erosita.mpe.mpg.de/edr/DataAnalysis/}}, which we use to perform data processing and reduction. The full details of eSASS are described in \cite{Brunner:2021hto}. Here, we only lay out the relevant procedures. 

\par We first use \texttt{evtool} from eSASS to combine the 4 observations of eFEDS into one stacked observation, with setting: \texttt{flag=0xC000F000} to remove unusually bright pixels, \texttt{gtitype=flaregti} to remove time intervals with flares, and \texttt{pattern=15} for photon reconstruction pattern (following the procedure in appendix A of~\cite{Brunner:2021hto}). We expect stacking these 4 observations would not influence the accuracy of sky background or DM signal model, since their locations are within a few degrees of each other. Using \texttt{srctool}, we then obtain 7 spectra with unit in ph/kev/s, one for each TM, as well as their corresponding Response Matrix Function (RMF) and Auxiliary Response Function (ARF) files. This allows us to account for 7 TM's slightly different internal background and response matrix. For \texttt{srctool}, we set parameter \texttt{exttype=point}. The target region is set to a box sized 20$^\circ$ by 10$^\circ$, centered at eFEDS location, to encompass the entirety of eFEDS field. The details of stacked eFEDS observation after processing are summarized in table~\ref{obsTM_table}. 

\begin{table}
\centering
\begin{tabularx}{0.45\textwidth}{YYYYY}
\hline
\hline
Telescope Module & Central RA & Central DEC & Combined, Cleaned Effective Exposure & Combined, Cleaned Sky area\\
 & (deg) & (deg) & (s) & (deg$^2$)\\
 \hline
TM1 & 136.00 & 1.50 & 360312 & 164.46\\
TM2 & 136.00 & 1.50 & 360583 & 163.52\\
TM3 & 136.00 & 1.50 & 357331 & 164.47\\
TM4 & 136.00 & 1.50 & 350152 & 164.49\\
TM5 & 136.00 & 1.50 & 301490 & 164.44\\
TM6 & 136.00 & 1.50 & 293827 & 164.42\\
TM7 & 136.00 & 1.50 & 360543 & 164.46\\
 \hline
 \hline
\end{tabularx}
\caption{Parameters of stacked eFEDS observation, encoded in file headers after \texttt{srctool} processing. The sky area is calculated as \texttt{REGION\_AREA*BACKSCAL}}
 \label{obsTM_table}
\end{table}%

\par To ensure employing the blank sky model is appropriate, we have investigated the significance of point sources in eFEDS. First, \texttt{ermask} is used to create a ``cheesemask" that blocks off all sources listed in eROSITA catalogue~\cite{Brunner:2021hto}. Next, we extract spectra of the same observation both with and without the ``cheesemask" applied. We have tested a particularly liberal masking that reduces exposure area by 60\%. The spectra extracted before and after masking provide similar goodness of fit parameter ($\chi^2/{\rm d.f.}$), for the same blank sky model for our spectral analysis. Therefore we decide to use the data without masking since it provides more effective exposure time and area. 

\par The final step of our data reduction is rebinning. Each spectra is rebinned to equal log width bins, specifically 200 bins per decade. In our binning scheme, the bin width is smaller than eROSITA energy resolution (FWHM), so that a Gaussian line would show up in multiple bins. For eROSITA, FWHM ranges from $\sim80$ eV at 1 keV, to $\sim160$ eV at 8 keV. After binning for individual TMs, each bin in our spectrum has roughly between 2000 to 6000 photons. We have tried using the default binning scheme given by eROSITA data, which is used by RMF and has 728 bins between 0.9 - 9 keV. In this binning scheme, there are 700 to 2500 photon counts in each bin. Between these two choices of binning, we see no significant difference in the best fit $\chi^2$/d.o.f. and DM decay rate constraint.

\par With the spectra from eFEDS in hand, we then move on to construct a blank sky models spectrum for eFEDS. 

\subsection{\label{sec:spectfit}Modeling eFEDS Blank Sky}

\par We perform the spectral fitting with XSPEC, version number 12.12.1~\cite{1996ASPC..101...17A}. Our null sky model has two parts, instrumental background and astrophysical background. 

\par The instrumental background is the dominant component above $\sim2.2$ keV. It is caused by various cosmic particles interacting with the instrument as well as other noises. The model is based on in-flight measurement taken during Cal-PV period, with camera filter wheel set to ``closed” position. Both the data and model of filter wheel closed observation are produced by eROSITA collaboration, and are made publicly available on EDR website~\footnote{\url{https://erosita.mpe.mpg.de/edr/eROSITAObservations/EDRFWC/}}. The model has four components to describe the continuum, a double broken power law and two additional power laws, all multiplied by an exponential modification. On top of the continuum there are 14 Gaussian lines to model emission from various internal elements~\cite{Whelan:2021osf}. During spectral fitting, we allow the three power law indices, the exponential modification factor, two break energies, and all 4 normalization parameters for continuum to float freely. The only exception is the index and normalization of the last power law for TM5 and TM7, which are frozen at 6 and 0, following FWC model provided by eROSITA EDR. For the instrument Gaussian lines, we fix the line widths, allow the line energy location to float within ranges preset by eROSITA FWC model, and allow the normalizations to float freely. We assume no correlation between free model parameters across different TMs and allow them to fit independently. 

\par We note that the major contribution to the continuum in instrumental background is the broken power law, while the two power law components matter very little, due to latter two's much lower normalization value than the former. This is a choice made by eROSITA collaboration when they produced the best fit instrumental background model. As understanding the background is beyond the scope of this study, we take an utilitarian approach and adopt the model components from EDR verbatim.

\par For astrophysical X-ray background, we consider (1) an absorbed \texttt{apec}~\cite{Smith:2001he} model with temperature $kT\sim0.75\rm\,keV$ for the Local Group emission, and (2) an absorbed power law with index around 1.46 accounting for unresolved extragalactic sources, sometimes called Cosmic X-ray background (CXB). This modeling choice is based on previous eROSITA eFEDS analysis~\cite{Liu:2021ewv}, which itself is based on XMM-Newton model in soft X-ray range~\cite{Kuntz_2000, Snowden:2007jg, Bulbul:2011ku}. In the two astrophysical components, all normalization parameters are free to vary. The temperature of the hot galactic halo \texttt{apec} model is allowed to vary within range of 0.5-1 keV. We use model \texttt{tbabs}~\cite{Wilms:2000ez} for intergalactic medium absorption, with he initial fit value of hydrogen column density value at eFEDS target location set to 4.4e20 cm$^{-2}$ and initial fit value of abundance set to solar, adopting abundance table from~\cite{Wilms:2000ez}. The calculation of hydrogen column density value is provided by UK Swift Science Data Centre~\cite{Willingale:2013tia}. To account for uncertainty from modeling, we allow all parameters to float, and assume there is no correlation between the parameters for each TM. The best fit parameters of our blank sky model roughly agree with the numbers reported in eROSITA diffuse sky study~\cite{Ponti:2022nix}. See the appendix table~\ref{all+model_param_table} for the exact values of our best fit parameters and their allowed ranges. We also test the impact of fixing CXB power law index to values reported in previous literature. We find that, the choices of different CXB power law indices change our DM constraint result by less than factor of 3 in most energies. See appendix~\ref{sec:plawind} and figure~\ref{fig:3plaw} within for further discussion.

\par There is another unabsorbed \texttt{apec} model component commonly present in previous astrophysical X-ray background modeling, meant to model Local Hot Bubble~\cite{Snowden:2007jg}. This \texttt{apec} peaks at around $kT\sim0.2$ - $0.25$ keV, which is well below the $0.9$ - $9$ keV energy range where we fit the spectrum and search for DM. We have tried fitting eFEDS spectrum with this unabsorbed apec component included alongside the absorbed apec and power law. The best fit $\chi^2$ and the DM search constraint show no noticeable change except near the low energy edge. Therefore, we conclude it is safe to exclude this component in our blank sky modeling.

\par Figure~\ref{fig:fit} shows data and fitted model for TM1, as a general representation of all 7 TM (see appendix figure~\ref{fig:bestfitTMALL} for spectra and best fit models of all 7 TMs). Figure~\ref{fig:fit} also shows a fitted model with fiducial DM signal (see section~\ref{sec:dm_analysis}). Besides the number of photon counts in each bin giving a statistical error at $\sim1.3-2\%$, we also add a systematic uncertainty at 1.5\% to account for the crudeness of our model and data processing. The reasons for adding systematic error include: dim X-ray sources not removed in the observation region, the slight variation in each patch of sky, and the imperfect model originally from XMM-Newton. The value 1.5\% is chosen such that our model spectrum for each TM has $\chi^2/$d.o.f.~$\sim1$. 

\begin{figure*}

    \centering
    \includegraphics[width=\linewidth]{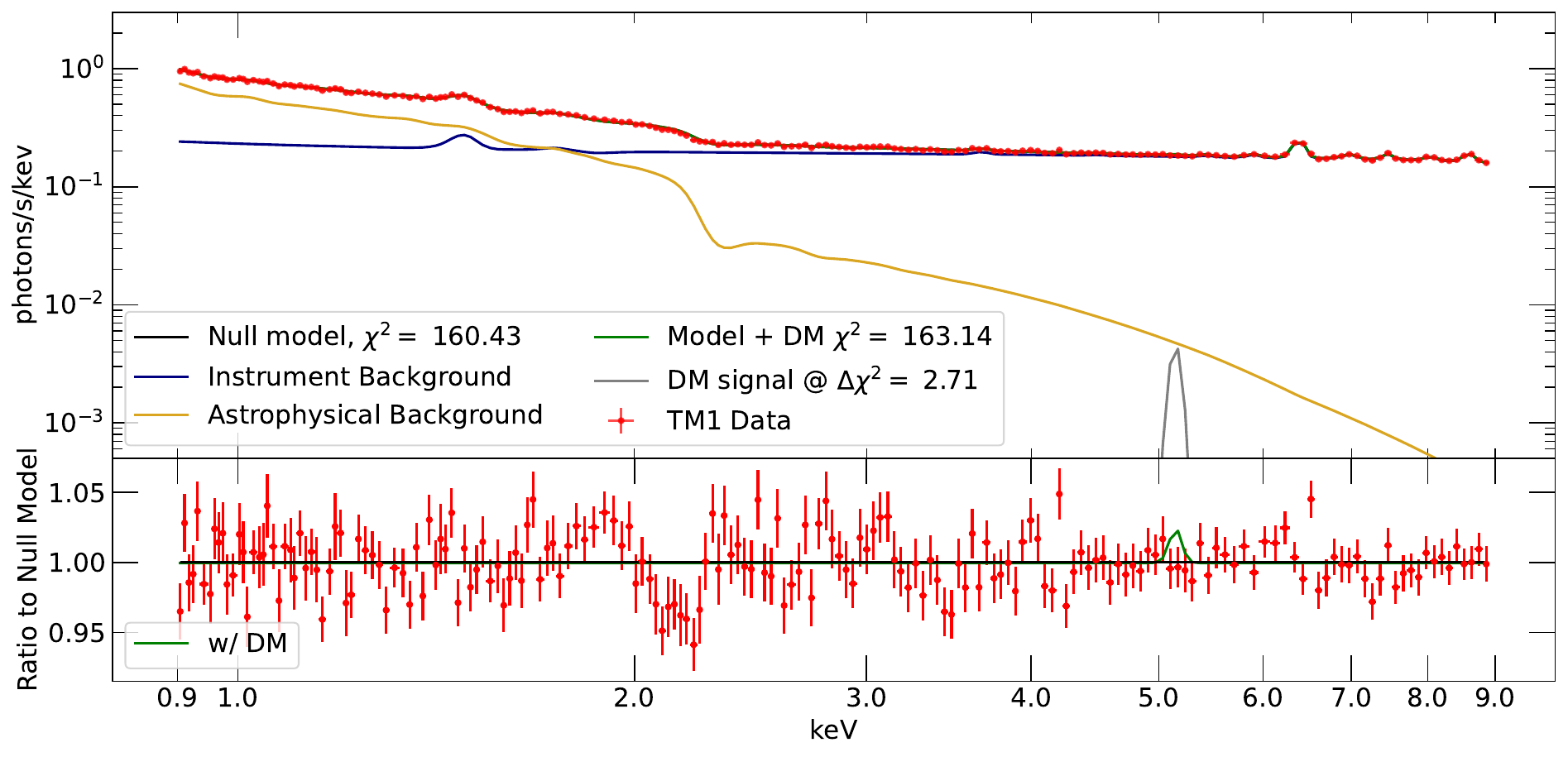}
    \caption{Data and model spectra for eROSITA eFEDS observation, obtained by TM1. The model spectrum consists of instrumental background (blue) and astrophysical background (yellow). There are 191 bins and 20 free parameters. Error bars correspond to $\pm 1\sigma$. This figure also shows a fiducial DM signal (grey) and a fitted model with this signal (green), to demonstrate our unexplained X-ray line search procedure (see section \ref{sec:dmsignal}). }
    \label{fig:fit}

\end{figure*}

\section{\label{sec:dm_analysis}Dark Matter Analysis}

\subsection{\label{sec:linesearch} eFEDS Line Search}

\par To search for a possible signature in eFEDS from DM decay, we fit the spectrum with a mock signal on top of our null model. The mock signal has a Gaussian shape, with normalization determined by free parameter decay rate $\Gamma$ and fixed parameter $m_\chi$, see equations in section~\ref{sec:dmsignal}. Regarding DM Gaussian line width, it is important to consider the potential Doppler shift resulted from DM velocity dispersion in Milky Way halo. Recent works find that the Doppler shift from velocity dispersion will result in DM line broadening $\Delta E/E\lesssim0.12\%$~\cite{Speckhard:2015eva, Powell:2016zbo, Zhong:2020wre, Dessert:2023vyl, Lovell:2024qwb}. This level of broadening is far lower than eROSITA energy resolution, which is between 2\%-8\% in the energy range we study~\cite{eROSITA:2020emt}. Therefore, in our analysis, we did not consider the potential line broadening from DM dispersion velocity, and set DM Gaussian line width to be 0 in our line search procedure. With DM Gaussian line parameters determined, we then fold the model through corresponding instrument responses functions (RMF and ARF) extracted with \texttt{srctool}, similar to all other models with astrophysical origin. The green and grey lines in figure~\ref{fig:fit} demonstrate of our fiducial signal, and a refitted model with said signal.

\par We scan for fiducial signal, through the spectral energy range 0.9-9 keV, in 200 equal logarithmic width energy steps, in accordance with our binning scheme. In each step all free parameters in the null model are allowed to vary. We take the conservative approach by assuming the line flux is degenerate with instrument and sky model. We combine the sensitivity of all 7 TMs by considering the parameter $X^2{\left(\Gamma\right)} = \sum_{i=1}^{7} \chi_i^2{\left(\Gamma\right)}$. A 5$\sigma$ detection of unexplained line would require $X^2{(\Gamma_{\rm det})} < X^2{(\Gamma = 0)} - 25$.

\par The line search procedure shows eFEDS observation is mostly consistent with our null model, with a notable exception of an excess at spectral energy 1.9 keV. Due to this excess, the best fit with DM indicates $\Delta X^2 \approx 100$ for $m_\chi=3.8\rm\, keV$ at decay rate $\Gamma = 10^{-28}\rm\,s^{-1}$. We do not believe this feature indicates a detection of DM or unexplained X-ray line, but rather mis-modeling of eROSITA's rapidly changing effective area in this energy range, also reported in other eFEDS studies (e.g., figure 6 of ~\cite{Liu:2021ewv}). eFEDS study on cluster Abell 3266 reported 10\% residual and suspected it's due to calibration uncertainty of the telescope~\cite{Sanders:2021oyk}. We will discuss the impact of these unexpected features on our DM constraint and our treatment in section~\ref{sec:dm_constraints}.

\par Next we turn to interpret the line search result in the context of galactic DM decay, and set upper limit on DM decay rate.

\subsection{\label{sec:dmsignal} Dark Matter Signal Modeling}

\par The general formula for DM decay flux at a certain sky location, in unit of $\rm ph\,cm^{-2}\,s^{-1}\,keV^{-1}$ is given by  
\begin{equation}\label{eq:D}
\frac{dF}{dE} = \frac{\Gamma}{4\pi m_\chi} \frac{dN}{dE} \cdot \int d\Omega \cdot \mathcal{D} \,. 
\end{equation}
In this equation, $E$ is photon energy, $\Gamma$ is the decay rate of DM particle, and $m_{\chi}$ is the mass of DM particle. $\frac{dN}{dE}$ is the spectrum of a single decay, which for this work we have taken to be a delta function, centered at half DM particle mass: $\delta(E-m_{\chi}/2)$. $\int d\Omega$ is the solid angle area of observation. $\mathcal{D}$ is D-factor, which calculates the column density of DM in field of view, given by equation:
\begin{equation}
    \mathcal{D} = \int_{\rm los} dl \rho\left(r\left(\psi,l\right)\right) 
\end{equation}
where $\rho(r)$ is the DM density profile, $r(\psi, l)=\left(R^2_\odot+l^2-2R_\odot l \cos\psi\right)^{1/2}$ is galactocentric radius, $l$ is line of sight distance, $\psi$ is pointing angle from Galactic Center, and $R_\odot=8.5$ kpc is the distance of solar system to Galactic Center. In our DM analysis, we evaluated the integral in eq. \ref{eq:D} by dividing the eFEDS observation region using \texttt{healpy}~\cite{Górski_2005, Zonca2019} with nside parameter = 128. This gives us 776 sub regions, with each being 0.21 deg$^2$. Then for each \texttt{healpy} pixel we calculated the D factor and summed over them.

\begin{figure}[t]

    \centering
    \includegraphics[width=\linewidth]{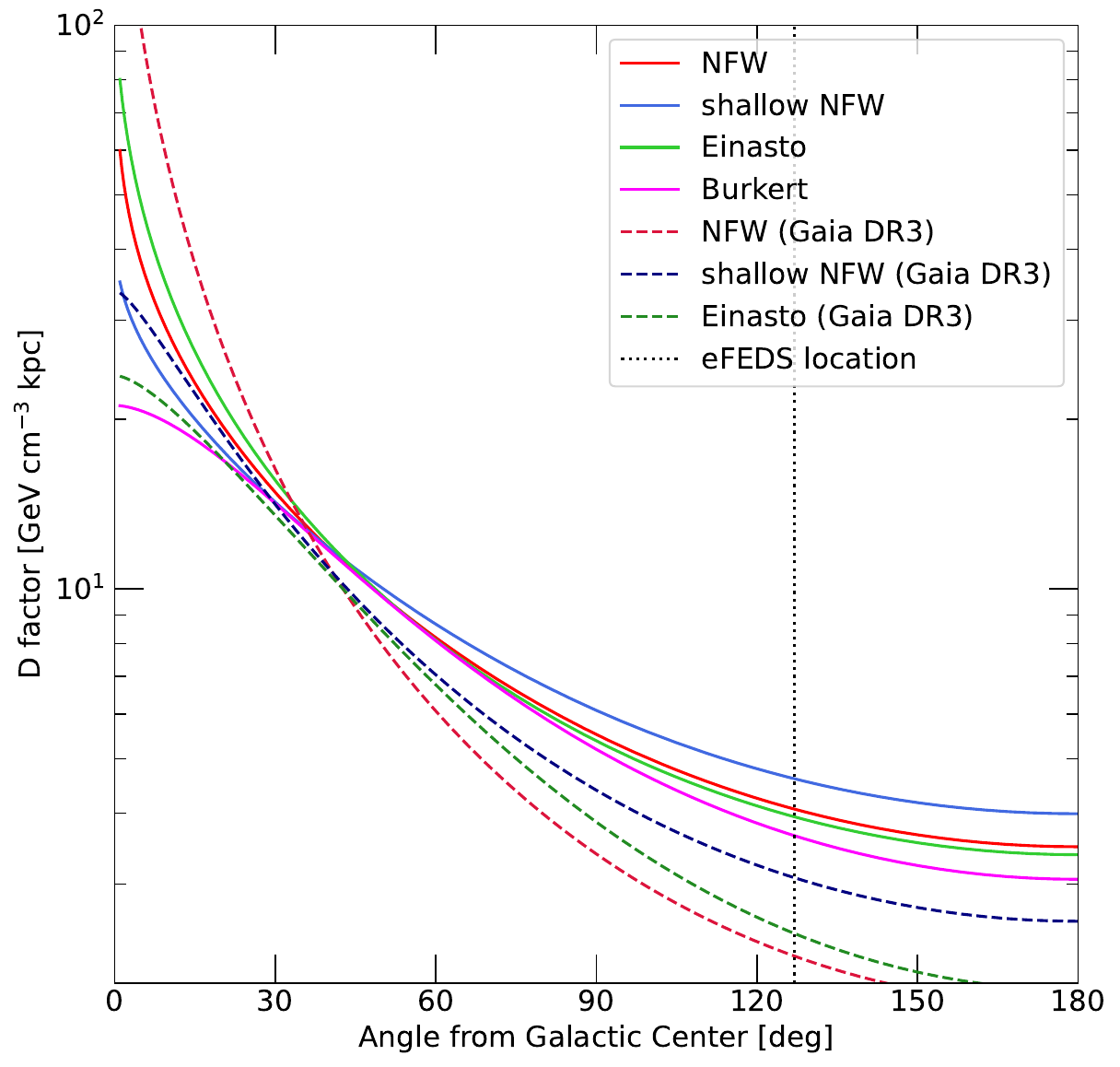}
    \caption{DM column density D-factor as a function of the observation angle to Galactic Center. DM profiles from different studies are considered in this plot: NFW, shallow NFW, Einasto, and Burkert. Dashed lines denote results calculated using Gaia DR3 studies~\cite{Lim:2023lss, Ou:2023adg}(see \ref{sec:dmsignal} for detailed explanation). The dotted line marks eFEDS central location's angle, demonstrating that in this analysis, there is marginal difference on D-factor in the choice of DM profile. }
    \label{fig:J_fac}

\end{figure}

\par Figure \ref{fig:J_fac} shows the D-factor of the density profiles we considered in this work, as a function of angle from galactic center. A few words should be said about the choice of DM density profile. A popular DM density profile in literature is Navarro-Frenk-White (NFW) profile \cite{Navarro:1996gj}, the generalized equation of NFW profile is described by:
\begin{equation}
\rho_{\chi}(r) = \rho_\odot\cdot\left(\frac{r}{R_\odot}\right)^{-\gamma}\left[\frac{1+\left(\frac{R_\odot}{R_s}\right)}{1+\left(\frac{r}{R_s}\right)}\right]^{3-\gamma}
\end{equation}
where $R_s = 20$ kpc is scaling distance~\cite{Hooper:2016ggc} and $\rho_\odot = 0.4$ GeV/cm$^3$ is DM density at solar neighborhood~\cite{deSalas:2019pee, deSalas:2020hbh, Sofue:2020rnl}. For Milky Way kinetic simulation, the original best fit parameter shows $\gamma=1$~\cite{Navarro:1996gj}. We denote this DM density profile with a cusp at the center as simply NFW. Numerical simulation with baryon added in shows the inner slope of DM could have a core~\cite{Calore:2015oya, Nesti:2013uwa}. To approximate this case, we use an NFW profile where the slope is shallow all the way to the center with $\gamma = 0.7$~\cite{Pato:2015dua} denoted as shallow NFW. 

\par We also consider Einasto profile~\cite{Diemand:2008in} described by
\begin{equation}
\rho_{\chi}(r) = \rho_\odot \cdot \exp\left[ -\frac{2}{\alpha}  \left( \left(\frac{r}{r_s}\right)^\alpha - \left(\frac{R_\odot}{r_s}\right)^\alpha \right) \right]    
\end{equation} 
with $R_s = 20$ kpc. The typical Einasto profile (EIN) has $\alpha = 0.17$. 

\par Lastly, we consider Burkert profile which was first introduced to model dwarf galaxy~\cite{Burkert:1995yz}
\begin{equation}
\rho_{\chi}(r) = \frac{\rho_0r_0^3}{(r+r_0)(r^2+r_0^2) }
\end{equation}
where $\rho_0$ is the characteristic density and $r_0$ is the scale length. We take the best fit parameters $\rho_0= 1.98\rm\,GeV\,cm^{-3}$ and $r_0=8\rm\,kpc$ estimated by a study using LAMOST and SDSS observations~\cite{Lin:2019yux}. 

\par Recently, two studies calculated best fit DM density profile from Gaia DR3~\cite{Lim:2023lss, Ou:2023adg}. However, the best fit DM density and D-factor obtained from these studies are noticeably lower than previous results, as shown in figure~\ref{fig:J_fac}. To ensure our DM analysis is consistent with previous DM studies and can be compared, we do not make use of these latest DM density profiles from Gaia DR3. For our analysis we take NFW profile as the most general description for DM and only report one sets of results. We note that depends on the specific DM density model, our constraint result could be modified by a factor of 2. 

\par We do not include the contribution from Extra Galactic (EG) DM decay into account. For EG DM, the Gaussian signal from a single decay is broadened to continuum as it gets redshifted. We find that the EG continuum is degenerate with the background continuum, and the line search procedure is still dominated by Gaussian. At eFEDS observation angle, EG DM photon counts should contribute as much as galactic DM, but only when integrated over all energy range.

\subsection{\label{sec:dm_constraints} Dark Matter Decay Rate Constraint and Sensitivity Analysis}

\par We obtain DM decay rate upper limit by using the same scanning and fitting procedure described in~\ref{sec:linesearch}. At each of the 200 steps, we scan through DM line normalization value, start with $\Gamma = 0$ and increasing it until the best fit model with DM line has 
\begin{equation}
    X^2{(\Gamma_{\max})} - X^2{(\Gamma_{\min})} =2.71.
    \label{chi2}
\end{equation}
In this equation, we look for $\Gamma_{\max}$, which corresponds to the 95\% one sided upper limit on DM decay flux. 

\par We use Monte Carlo (MC) procedure to examine the validity of our result. We generate 1000 mock spectra for each of the 7 TM based on their best fit model using \texttt{fakeit} function from XSPEC, assuming Poisson fluctuation. Each mock spectrum has the same exposure, response function, and effective area as the real data. We then pass all data sets through the same line search procedure. The number of tested masses was reduced from 200 to 100 to speed up the algorithm. Armed with 1000 simulated upper limits, we then constructed 68\% and 95\% upper limit intervals, corresponding to 1$\sigma$ and 2$\sigma$ confidence level respectively. 

\par
\begin{figure}

    \centering
    \includegraphics[width=\linewidth]{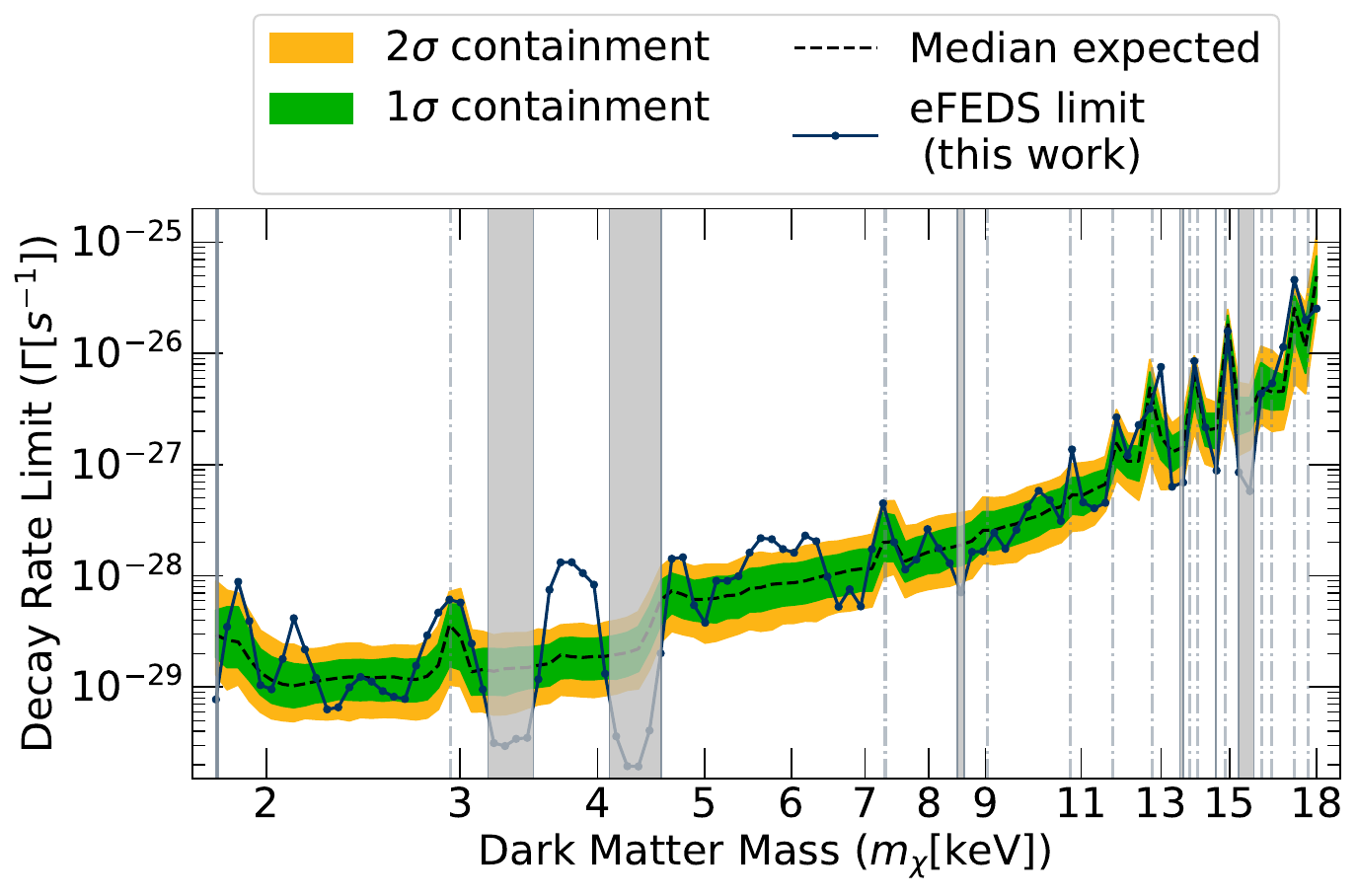}
    \caption{95\% DM decay rate upper limit obtained from eROSTIA eFEDS observation, assuming the decay product has a single photon. Green and gold bands here denote expected $1\sigma/2\sigma$ containment on decay rate under null hypothesis. The grey dash dotted lines mark the energies where our instrument model has a Gaussian line. Grey band regions mark energies where our limit is stronger than $2\sigma$ containment.}
    \label{fig:MC}

\end{figure}

 In figure~\ref{fig:MC}, we present our model independent DM decay rate constraint from eFEDS observation, as well as the statistical containment from MC procedure. Our decay rate limit is applicable to any DM model that has one photon in its decay product. To compare it to 2 photons decay DM models with, one needs to divide this limit by a factor of 2 (e.g., ALP limit in section~\ref{sec:ALP_limit}). 

\par Since our analysis conservatively assumes a potential DM signal and the Gaussian lines from instrument model are indistinguishable, our DM lifetime limit is weaker where the instrument background also has a Gaussian line. The places where the background model has a Gaussian component are marked on figure~\ref{fig:MC} with grey dash dotted lines.

\par Overall, most regions of our limit falls within 2$\sigma$ containment region. There are a few places where it goes beyond. As shown in the Monte Carlo analysis, our limit exceeds two sigma containment at 4-5 keV. The energy in photon spectrum corresponding to the problematic DM mass is 2-2.5 keV, which shows two “troughs” and a “bumps”. We think all of them are due to inaccuracies in calibration for effective area by eROSITA collaboration at this energy range, as discussed in section~\ref{sec:linesearch}. In particular, the limit exceeds 2$\sigma$ containment at around $m_\chi=3$-$6\rm\,keV$. At this energy range, we argue it reflects a modeling shortcomings in spectral energy $1.5$-$3\rm\,keV$. As shown in figure 9 and 10 of~\cite{eROSITA:2020emt}, the effective area, as well as grasp, of eROSITA detector drop off sharply between $2$-$3\rm\,keV$, followed by making a tiny bump. The quirks in our limit is a direct consequence of effective area calibration done by eROSITA collaboration. One recent eROSITA study on 3.5 keV line investigated the possibility of this feature coming from mis-modeling of gold absorption edge, but ultimately concluded gold edge alone is not enough to explain the residual~\cite{Villalba:2024uyh}. With an improved effective area model, we expect this deviation to disappear in future analysis.

\par Our procedure for line search and setting constraints would deviate from Monte Carlo expectation when the observed photon spectrum has deficit and excess from the model. When the observed photon spectrum has an excess over the model, our limits become weaker than expected, and thus is conservative, but our line search procedure still can rule out DM signal above the constraint. As it would not falsely rule out the presence of a signal due to the profiling procedure in our analysis, we keep the conservative limit in our results. On the other hand, since the deficits would cause the constraint to mistakenly rule out DM parameters, we choose to exclude the overly stringent limit below $2\sigma$ containment. This regions are marked on figure~\ref{fig:MC} as grey vertical bands.

\subsection{\label{sec:MC}Future eROSITA Survey Projection with Mock Data}

\par
\begin{figure}
    \centering
    \includegraphics[width=\linewidth]{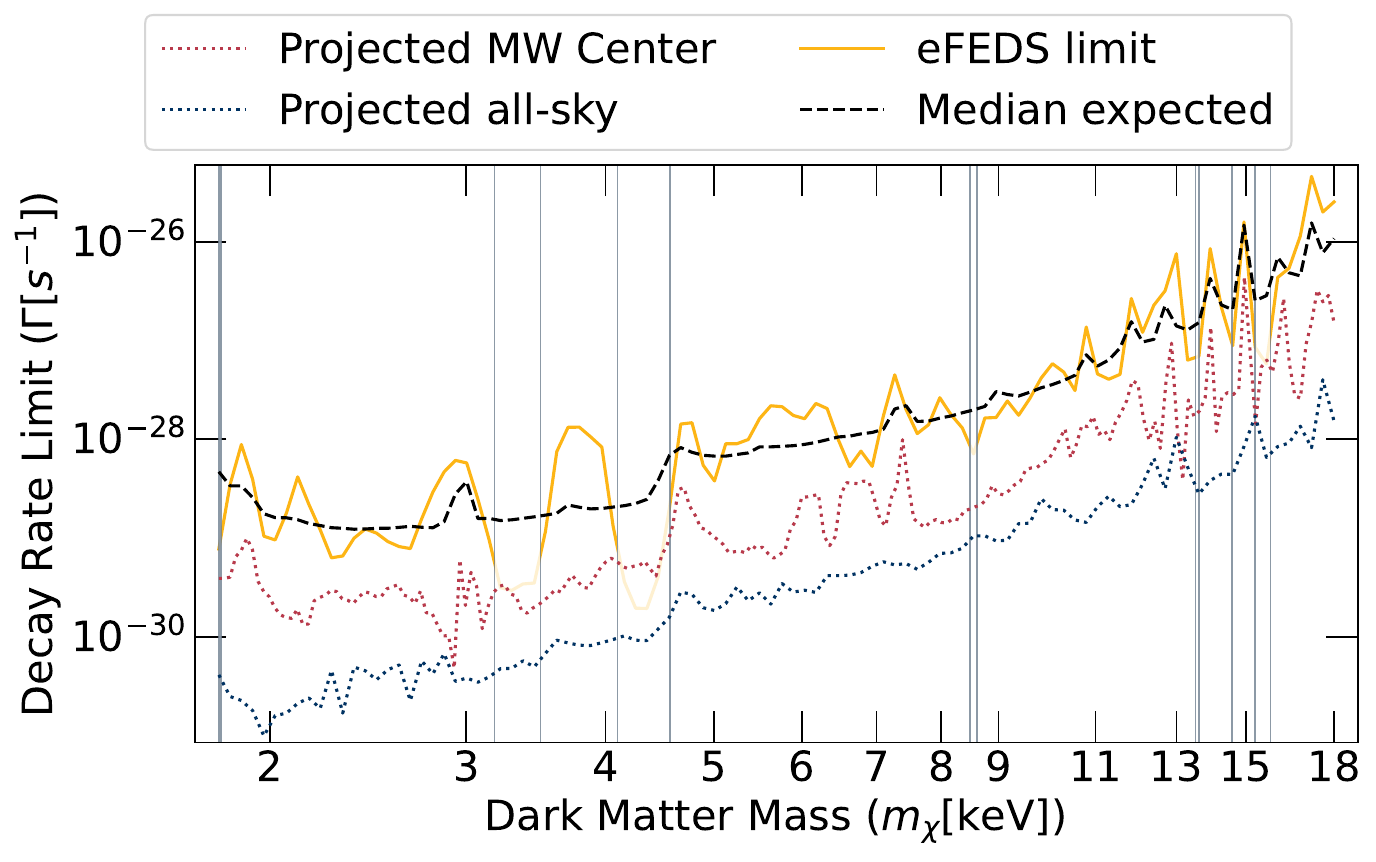}
    \caption{Projected DM decay rate sensitivity of eROSITA future MW center and all-sky survey, based on the eFEDS analysis completed in this work. eFEDS limit and median expected limit from MC procedure are shown for comparison. Similar to figure~\ref{fig:MC}, vertical bands mark the exclusion region to eFEDS limit.}
    \label{fig:proj}
\end{figure}

 Figure \ref{fig:proj} shows our projected future eROSITA survey sensitivity to DM decay based our result from eFEDS. We also show the limit obtained from eFEDS itself and from mock data on figure \ref{fig:proj} for comparison.

\par We obtain the projected sensitivity using mock spectra generated for each pixel on a sky map. Specifically, we generate a sky map with 300 pixels, using \texttt{healpy}~\cite{Górski_2005, Zonca2019} with parameter Nside=5. Each pixel is 126 deg$^2$ large, roughly the same order of magnitude as eFEDS observation at 163 deg$^2$. We exclude the center pixel in this sky map situated at $(l,b) = (0^\circ, 0^\circ)$ for our analysis, since in this region DM density profile has the largest uncertainty.

\par For projected Galactic Center (GC) sensitivity, we use the pixel on the \texttt{healpy} sky map that does not cover $(l,b) = (0^\circ, 0^\circ)$, instead situated at $(l,b) = (9^\circ, 7^\circ)$ making its galactocentric angle $11.8^\circ$. We generate 7 mock spectra using the XSPEC's \texttt{fakeit} function, corresponding to 7 TMs of eROSITA. We assume eROSITA will observe this region with the same exposure as eFEDS at 360 ks. The mock spectra follow the same best fit model and parameters for each TM used in our main eFEDS analysis, and assumes the same systematic error. We pass the mock spectra through the same line search and DM analysis procedure as the main eFEDS analysis described in section~\ref{sec:dm_constraints}. The obtained limit is our projected GC sensitivity, with the D-factor calculated using the pixel's center point. 

\par For all-sky survey sensitivity projection, we take the sky map with 299 pixels (after excluding the center pixel), as the total sky area observed by eROSITA suitable for DM analysis. Similar to the GC survey pixel, we assume in the future eROSITA will complete all-sky survey by observing every pixel on the map with the same exposure time as eFEDS at 360 ks. Next, separately for each pixel, we generate 7 mock spectra using the XSPEC's \texttt{fakeit} function, making a total of 2093 mock spectra representing all the data from eFEDS's all-sky survey. Finally, we pass these mock spectra through a slightly modified DM decay rate constraint procedure. For all-sky sensitivity, we sum over pixel index, and find the projected limit by: 
\begin{equation}
\sum\limits^{299}_{j=1} X_j^2(\Gamma_{\rm\max}) - \sum\limits^{299}_{j=1} X_j^2(\Gamma_{\rm\min}) = 2.71. 
\end{equation}
In this equation, $X_j^2$ corresponds to the summed best fit $\chi^2$ from 7 TM mock data, and $j$ corresponds to the pixel index going from 1 to 299. The $X_j^2(\Gamma)$ profile is calculated individually for each pixel with the D-factor calculated using the center point of the pixel. 

\par We note that our procedure, for both GC and all-sky projection, ignores the background effect coming from Galactic disk and point sources. For DM line search performed in this work, we expect the impact of continuum background component to be relatively minor. Line background on the other hand, could be important due to signal background confusion. Thus the projected sensitivities reported in figure~\ref{fig:proj} represent slightly optimistic cases. We defer a detailed calculation taken all background component into account to future eROSITA DM analysis. Taken at face value, eROSITA's GC sensitivity improves from eFEDS observation by 10, which is the ratio of D-factor. The all-sky sensitivity on the other hand roughly improves from eFEDS by ratio of all-sky average D-factor times $\sqrt{300}$, as a result of increase in statistics. We will discuss the specific implications of these projections in the next sections.

\subsection{\label{sec:nu_limit}Sterile Neutrino Dark Matter Constraints}

\par In this section we use decay rate limits to constrain the parameter space of sterile neutrino DM. Sterile neutrino as DM is a well motivated and popular theory that has garnered many investigations. With a non-zero mixing angle, sterile neutrino DM can decay into an active neutrino and a photon with energy $E_\gamma = m_\chi/2$, and thus detectable through astrophysical observation. Certain frameworks with more sterile neutrinos, for example, neutrino minimal Standard Model ($\nu$MSM), can also explain baryogenesis, neutrino mass, and matter/antimatter asymmetry besides DM~\cite{Bringmann:2022aim, Asaka:2005an, Asaka:2006nq, Canetti:2012kh, Canetti:2012vf}.

\par Sterile neutrino $\nu_s$ decay rate depends on the mass $m_\chi$ and the mixing angle $\sin^2(2\theta)$, given by formula~\cite{Abazajian:2001nj, Abazajian:2001vt, Shrock:1974nd, Pal:1981rm}:
\begin{equation}\label{eq:decay_rate}
    \Gamma_{\nu_s\rightarrow \nu \gamma} = 1.38\times 10^{-32}\,{\rm s^{-1}}\left( \frac{\sin^{2}2\theta}{10^{-10}} \right)
    \left( \frac{m_{\chi}}{\rm keV} \right)^{5}.
\end{equation}

\par
\begin{figure}[t]

  \centering
  \includegraphics[width=\linewidth]{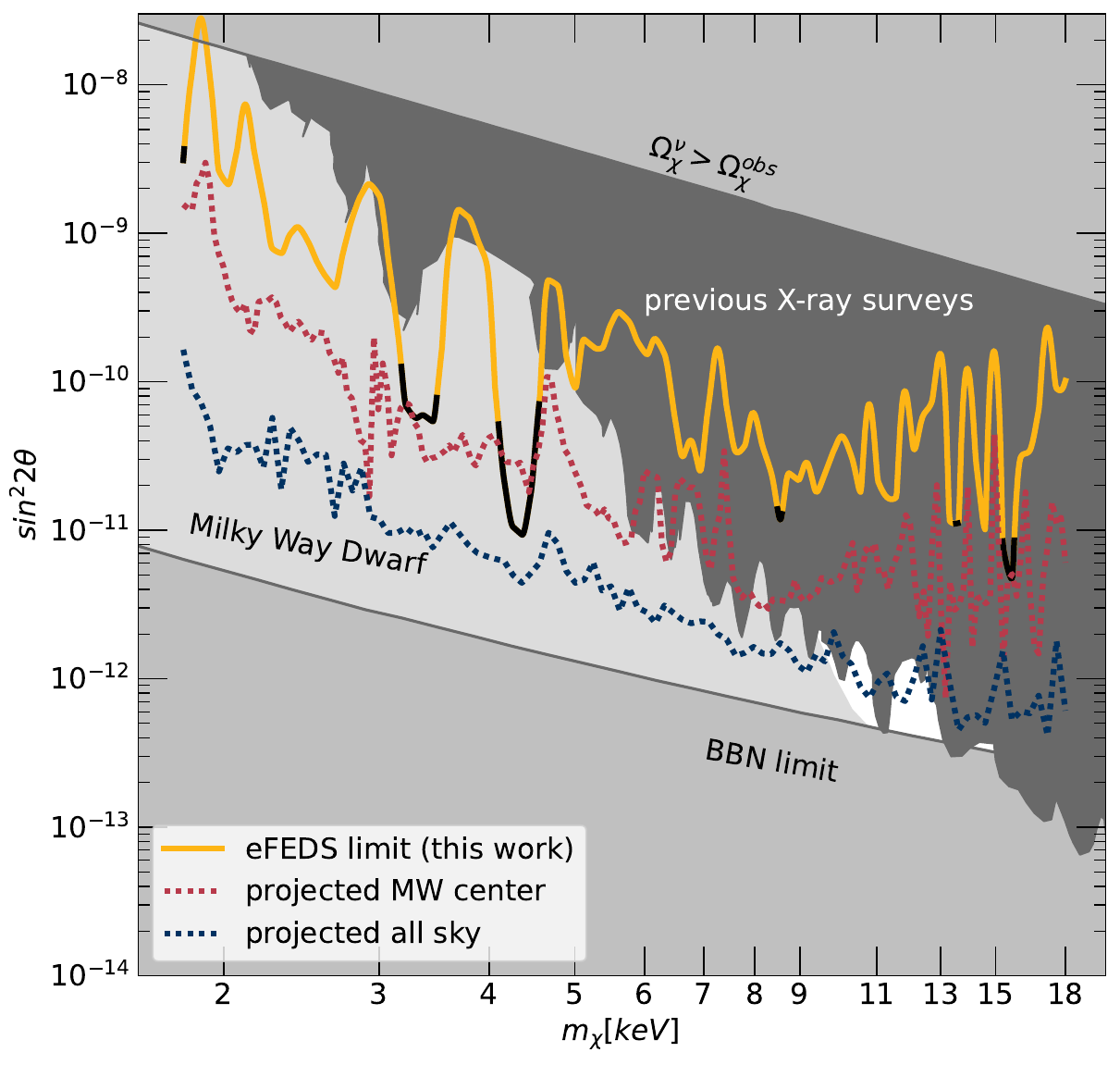}
  \caption{Current constraints on $\nu$MSM parameter space, including the one produced by eFEDS (gold line, this work), and projected eROSITA constraint from future eROSITA Milky Way center (red) and all-sky surveys (blue). The hatched area denotes regions where eROSITA limit surpasses all previous astrophysical X-ray constraints. Black lines denote where the limit exceeds $2\sigma$ lower containment. The previous X-ray constraints shown here include the ones derived from Chandra M31 observation~\cite{Horiuchi:2013noa}, XMM-Newton all sky~\cite{Dessert:2018qih, Foster_2021}, NuSTAR~\cite{Neronov:2016wdd, Perez:2016tcq, Ng:2019gch, Roach:2019ctw, Roach_2023}, and Suzaku~\cite{Loewenstein:2008yi}. The BBN lower bound on $\sin^2 2\theta$ is calculated from \textsc{sterile-dm}~\cite{Venumadhav:2015pla}, given the lepton asymmetry consideration from BBN, $L_6 < 2600$~\cite{Boyarsky:2009ix, Laine:2008pg, Dolgov:2000ew}. Milky Way dwarf galaxy limit is derived from SDSS data, taken from~\cite{Cherry:2017dwu} (See also Ref.~\cite{Dekker:2021scf} and text for discussion). The last two constraints are specific to $\nu$MSM and can change if the production mechanism is different. }
  \label{fig:sin2t}
\end{figure}

 In figure~\ref{fig:sin2t}, we show sterile neutrino DM constraint derived from eFEDS, projected future eROSITA sensitivities, and all the existing constraints. In eROSITA energy space, relevant previous X-ray surveys include: Chandra~\cite{Horiuchi:2013noa, Sicilian:2020glg}, XMM-Newton~\cite{Watson:2006qb, Dessert:2018qih, Foster_2021}, NuSTAR~\cite{Neronov:2016wdd, Perez:2016tcq, Ng:2019gch, Roach:2019ctw, Roach_2023}, and Suzaku~\cite{Loewenstein:2008yi}. The limit obtained by eFEDS in this work is the strongest X-ray limit in $\sim1.8$-$4.8\rm\,keV$ (see also figure~\ref{fig:gagg} in section~\ref{sec:ALP_limit}). This is due to the long exposure time of eFEDS, which, after co-adding up all 7 TMs, gives $\sim2\,\rm Ms$ of exposure time. Our limits are roughly consistent with previously eROSITA projections~\cite{Dekker:2021bos}, and SRG sterile neutrino search with correlation function~\cite{Barinov:2022kfp} 

\par Sterile neutrino DM can be produced through mixing with active neutrinos in the early universe with Shi-Fuller mechanism~\cite{Shi:1998km}. This mechanism proposed that in the presence of large lepton asymmetry, even a small mixing angle could be enhanced by extra matter potential, therefore producing the amount of DM needed. In the special case of zero lepton asymmetry, and thus the production is non-resonant, it is known as Dodelson-Widrow mechanism~\cite{Dodelson:1993je}. Assuming sterile neutrino DM is produced by Shi-Fuller mechanism, one also needs to consider constraint from big bang nucleosynthesis (BBN), on the amount of lepton asymmetry in the universe~\cite{Canetti:2012kh}. We show the lower bound on $\sin^2 2\theta$~\cite{Boyarsky:2009ix, Laine:2008pg, Dolgov:2000ew}, which is calculated from \textsc{sterile-dm}~\cite{Venumadhav:2015pla}, given the lepton asymmetry consideration from BBN, $L_6 < 2600$.

\par 
In structure formation, the lighter DM models, or ``hotter" ones tend to suppress small scale structures, thus provides another indirect constraint~\cite{Tremaine:1979we, Dalcanton:2000hn,Cherry:2017dwu,Alvey:2020xsk, Schneider:2016uqi, Dekker:2021bos, Dekker:2021scf, DES:2020fxi}. Thus, we also show the conservative Milky Way dwarf galaxy limit derived from SDSS data, taken from~\cite{Cherry:2017dwu}. This constraint result relies on several assumptions, such as the total mass of milky way galaxy, the anisotropy in satellite galaxy distribution, etc. We note that recently analyses have generally produce stronger limits, except for taking the most conservative assumptions~\cite{Dekker:2021scf}. The canonical limit reported in Ref.~\cite{Dekker:2021scf} already covers all available parameter space. 

\par Comparing to existing limits, our limit from eFEDS cannot touch the remaining parameter space at 9-14 keV region. With Galactic central region's denser DM profile, eROSITA constraint be improved by a factor of 3, which would surpass XMM-Newton's result $m_\chi\lesssim7\rm\,keV$. With eROSITA all-sky sensitivity, the result can surpass NuSTAR's constraint and become the leading X-ray constraint below $\lesssim$ 13 keV. There will also be possibilities open up for studying decay signal from Dwarf galaxies~\cite{Ando:2021fhj}. By then, eROSITA will contribute to closing out the remaining spaces in $\nu$MSM. However, considering sterile neutrino DM produced through other mechanisms can open up new parameter space available for X-ray searches~\cite{Bringmann:2022aim, Alonso-Alvarez:2021pgy, Holst:2023hff, An:2023mkf}.

\subsection{\label{sec:ALP_limit}Axion Like Particle Dark Matter Constraints}

\par Axion-like particles (ALP) is another interesting DM candidate, which we will focus on in this section. Unlike Axion, ALP's coupling strength and mass are independent, and is a Nambu-Goldstone boson that may arise from spontaneous breaking of U(1) symmetry. ALP can couple to photon, allowing it to decay into two photons $a \rightarrow \gamma \gamma$. Each decay photon has the energy of $\frac{m_{a}}{2}$. ALP coupling to photon is described by~\cite{Sikivie:1983ip, Raffelt:1987im}: 

\begin{equation}
\begin{aligned}
\mathcal{L} \supset \frac{1}{4} g_{a \gamma \gamma} a F_{\mu \nu} \tilde{F}^{\mu \nu},
\end{aligned}
\end{equation}
where $a$ is pseudoscalar ALP field, $F_{\mu \nu}$ is electromagnetic stress tensor and $\tilde{F}^{\mu \nu}$ is its dual tensor, and $g_{a \gamma \gamma}$ is the coupling constant.

\par ALP's decay rate into photons can be calculated by~\cite{Pospelov:2008jk, ParticleDataGroup:2022pth, Higaki:2014zua}:
 
\begin{equation}
\begin{aligned}
\Gamma_{a \rightarrow \gamma \gamma} \simeq 5 \times 10^{-29}\left(\frac{m_a}{7 \mathrm{keV}}\right)^3\left(\frac{f_a}{5 \times 10^{14} \mathrm{GeV}}\right)^{-2} \mathrm{~s}^{-1},
\end{aligned}
\end{equation}%
where $m_a$ is the mass of ALP. $f_a$ is the decay constant, related to coupling constant $g_{a \gamma \gamma}$ by
\begin{equation}
f_a \equiv \frac{\alpha C_{a \gamma \gamma}}{2 \pi g_{a \gamma \gamma}}.
\end{equation}
Note that $C_{a\gamma\gamma}$ can take other values too~\cite{ParticleDataGroup:2022pth, Kim:1998va, DiLuzio:2016sbl, DiLuzio:2017pfr, DiLuzio:2020wdo}. We use $C_{a \gamma \gamma}=8/3-1.92\approx 0.75$ here, as in grand unification DFSZ model~\cite{Irastorza:2018dyq, DINE1981199, Zhitnitsky:1980tq}. For example,~\cite{AxionLimits}'s X-ray ALP limit uses $C_{a\gamma\gamma}=-1.92$, following KSVZ model~\cite{Kim:1979if, Shifman:1979if}. We caution the reader that the limit on $g_{a\gamma\gamma}$ can change, depending on the ALP model's $C_{a\gamma\gamma}$ 

\par To account for the production of two photons, we divide our one photon decay rate upper limit in section~\ref{sec:dm_constraints} by 2. We also convert existing X-ray bounds on sterile neutrino mixing angle to flux, and then convert to bounds on ALP coupling.

\par
\begin{figure}
  \centering
  \includegraphics[width=\linewidth]{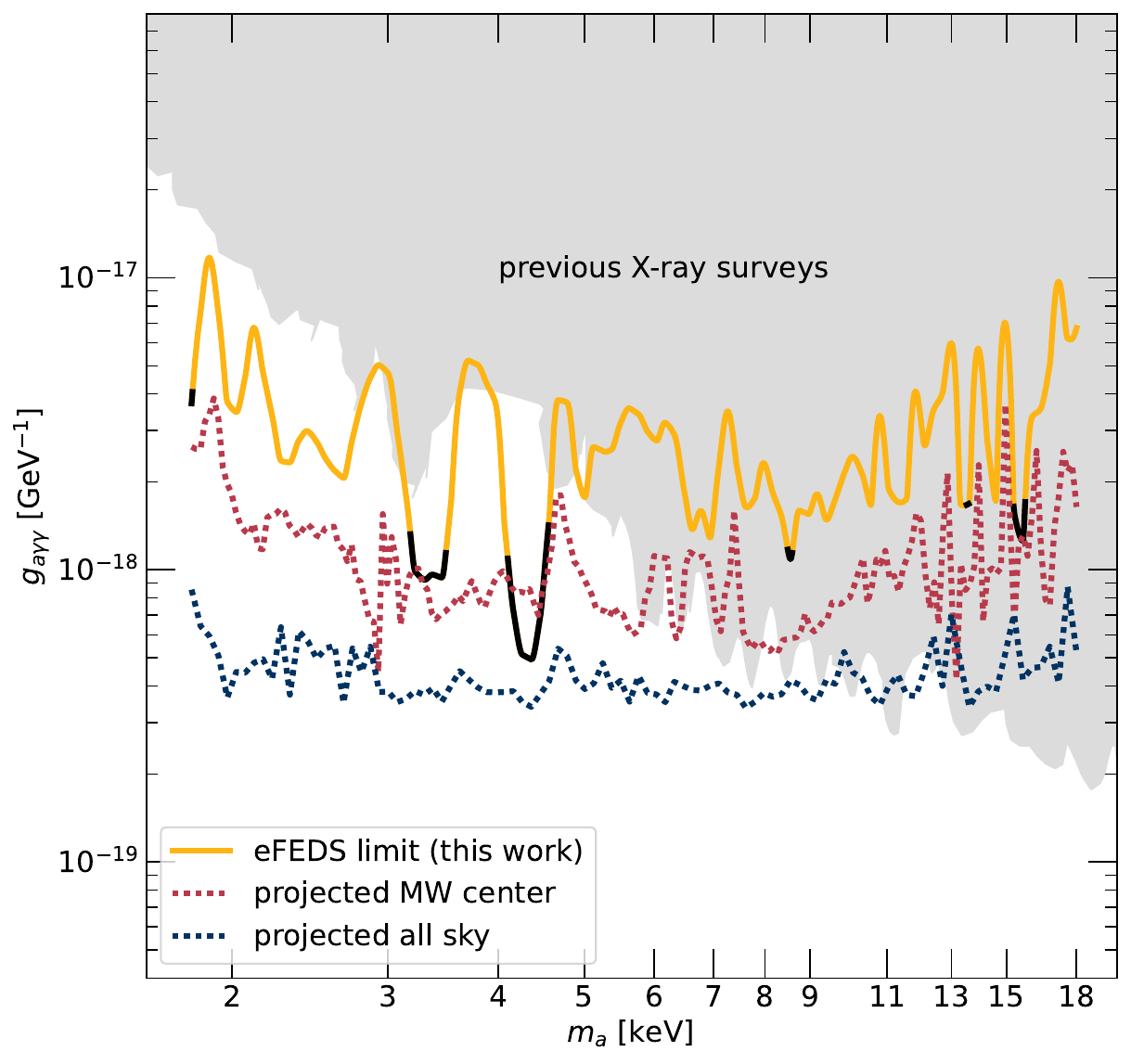}
  \caption{Current and projected eROSITA ALP limits, on top of previous X-ray survey results. Same as figure~\ref{fig:sin2t}, black lines indicate regions where eFEDS limit exceeds $2\sigma$ lower containment. The previous X-ray constraints shown here include the ones derived from Chandra M31 observation~\cite{Horiuchi:2013noa}, XMM-Newton all sky~\cite{Dessert:2018qih, Foster_2021}, NuSTAR~\cite{Neronov:2016wdd, Perez:2016tcq, Ng:2019gch, Roach:2019ctw, Roach_2023}, and Suzaku~\cite{Loewenstein:2008yi}.}
  \label{fig:gagg}
\end{figure}

 We show our ALP parameter space constraint in figure~\ref{fig:gagg}. As a non-thermal DM candidate, it does not need to conform to the structure formation constraints. As such, we only compare our result to previous X-ray surveys. eFEDS limit already exceeds previous constraints in energy range below 5 keV, set by Chandra M31 observation~\cite{Horiuchi:2013noa}. In the future, with Galactic Center observation, the limit set by eROSITA can become the leading constraints up to 7 keV.  

\section{\label{sec:conclusions}Conclusions and Outlook}

\par In this paper we use eFEDS X-ray observation, the largest and deepest observation set from eROSITA early data release, to search for and constrain decaying dark matter models. We combine 4 observations from eFEDS as one continuous observation, but treat data set from 7 Telescope Modules separately and only co-added them statistically. We construct sky and instrument model in energy range between 0.9-9 keV to avoid the light leak issue discovered in eROSITA observation.

\par Finding no unexplained X-ray line in our modeling, we set an upper limit on DM to photon decay rate in spectral range 0.9-9 keV, corresponding to DM mass 1.8-18 keV. We show that eFEDS limit can surpass all previous X-ray constraints for DM mass $\lesssim5$ keV. We also show with projected Milky way center and all-sky survey, eROSITA can provide the strongest constraint in energy range below 6.5 keV and 14 keV, respectively.

\par Using Monte Carlo method, we test the validity of our result. We have demonstrated the result lies within 2$\sigma$ containment in most of the energy range. We believe the overly strong limit in 4-6 keV range is due to eROSITA's rapid changing and uneven grasp in 2-3 keV. As more analyses are performed to perfect the effective area model of eROSITA, this quirk should disappear in future DM search constraints.

\par In the near future, with massive improvement in data size provided by eROSITA full data release, coupled with techniques presented in this work, we believe it can be a powerful survey to constrain decaying dark matter particles.  

\section*{\label{sec:acknowledgements}Acknowledgments}

\par This work is based on data from eROSITA, the soft X-ray instrument aboard SRG, a joint Russian-German science mission supported by the Russian Space Agency (Roskosmos), in the interests of the Russian Academy of Sciences represented by its Space Research Institute (IKI), and the Deutsches Zentrum für Luft- und Raumfahrt (DLR). The SRG spacecraft was built by Lavochkin Association (NPOL) and its subcontractors, and is operated by NPOL with support from the Max Planck Institute for Extraterrestrial Physics (MPE). The development and construction of the eROSITA X-ray instrument was led by MPE, with contributions from the Dr. Karl Remeis Observatory Bamberg \& ECAP (FAU Erlangen-Nuernberg), the University of Hamburg Observatory, the Leibniz Institute for Astrophysics Potsdam (AIP), and the Institute for Astronomy and Astrophysics of the University of Tübingen, with the support of DLR and the Max Planck Society. The Argelander Institute for Astronomy of the University of Bonn and the Ludwig Maximilians Universität Munich also participated in the science preparation for eROSITA. The eROSITA data shown here were processed using the eSASS software system developed by the German eROSITA consortium. 

\par The computational aspects of this work made extensive use of the following packages: the \texttt{SciPy} ecosystem~\cite{scipy}, particularly \texttt{Matplotlib} and \texttt{NumPy}; and \texttt{Astropy}, a community-developed core Python package for Astronomy \citep{astropy:2013, astropy:2018}; the package \texttt{multiprocess} \cite{2012arXiv1202.1056M, McKerns_Aivazis_2010}. Some of the results in this paper have been derived using the \texttt{healpy} and
\texttt{HEALPix} package~\cite{Górski_2005, Zonca2019}. This research has made use of data and/or software provided by the High Energy Astrophysics Science Archive Research Center (HEASARC), which is a service of the Astrophysics Science Division at NASA/GSFC and the High Energy Astrophysics Division of the Smithsonian Astrophysical Observatory. 

\par The works of CF, KCYN and QSL are supported by KCYN is supported by Croucher foundation, RGC grants (24302721, 14305822, 14308023), NSFC/RGC joint research scheme (N CUHK456/22), and NSFC grant 12322517. This work is supported in part by the National Natural Science Foundation of China under grant No. 12342502.

\bibliography{bib.bib}
\appendix
\counterwithin{figure}{section}

\section{\label{sec:individual_TM}Observed Data and Fitted Model of Each Individual Telescope Module}
\par
\begin{figure*}[hbt]
    \includegraphics[width=0.9\linewidth]{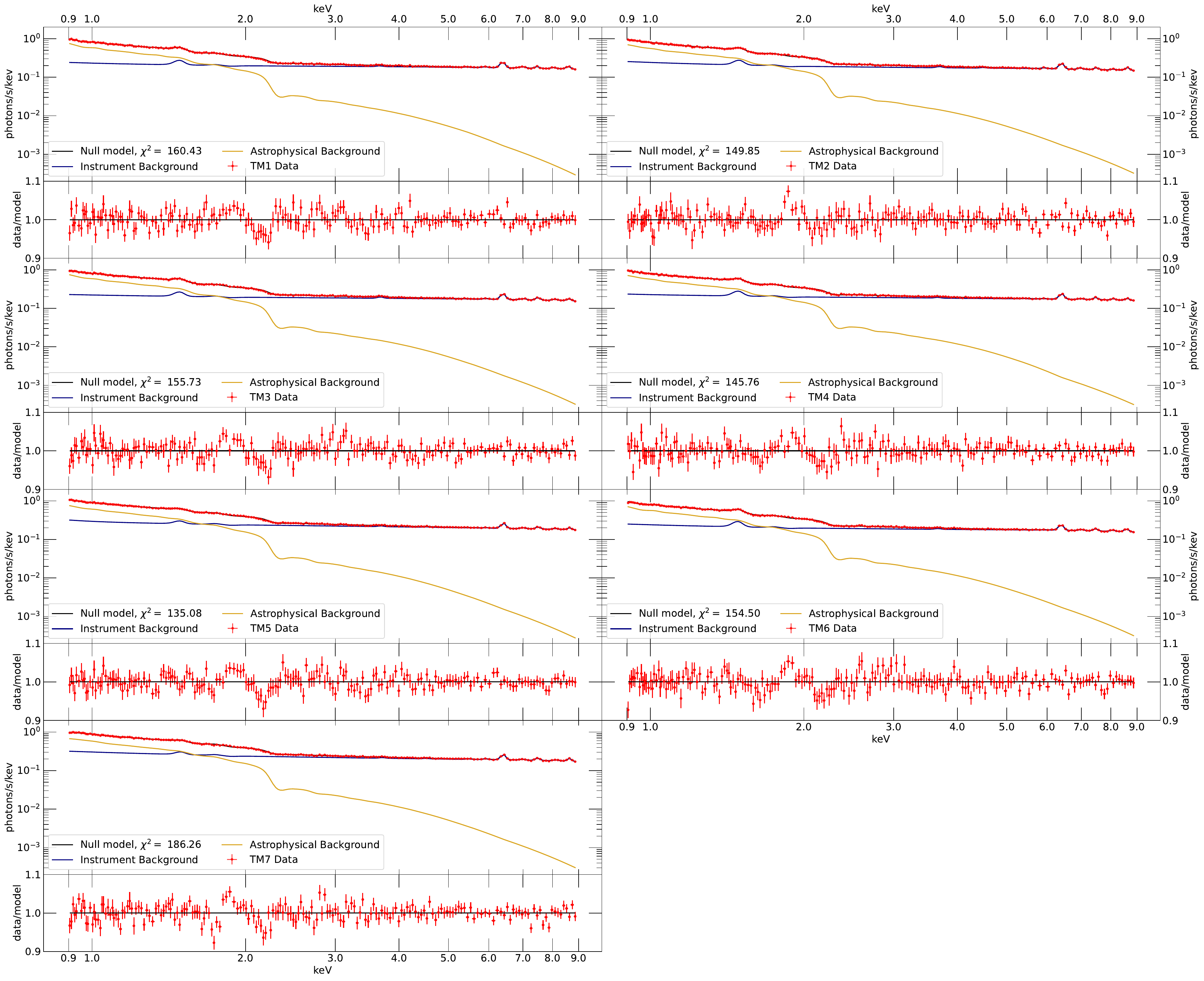}
    \caption{TM1-7 data and best fit model. Best fit $\chi^2$ is shown in figure legend.}
    \label{fig:bestfitTMALL}
\end{figure*}

 In the main text, we report the fitting procedure for our blank sky and DM model. We only show the best fit result for TM1 as an example in figure~\ref{fig:fit}. Here, we report the best fit results for all 7 TMs in figure~\ref{fig:bestfitTMALL} and all best fit parameter values in table~\ref{all+model_param_table}. There are 145 degrees of freedom for TMs not affected by light leak, and 147 d.o.f. for TM5 and TM7.
\begin{table*}[h]
\resizebox{\textwidth}{!}{%
\begin{tabular}{ccc|ccccccc|c}
\hline
\multicolumn{1}{c|}{Origin} & \multicolumn{1}{c|}{Model name} & Model Param. (unit) & \multicolumn{7}{c|}{Telescope Module} & Freeze Status (Range) \\ \hline
 &  &  & \multicolumn{1}{c|}{1} & \multicolumn{1}{c|}{2} & \multicolumn{1}{c|}{3} & \multicolumn{1}{c|}{4} & \multicolumn{1}{c|}{5} & \multicolumn{1}{c|}{6} & 7 &  \\ \hline
\multicolumn{1}{c|}{\multirow{7}{*}{Astrophysical}} & \multicolumn{1}{c|}{\multirow{4}{*}{apec}} & Abund (Solar) & \multicolumn{1}{c|}{1} & \multicolumn{1}{c|}{1} & \multicolumn{1}{c|}{0.99} & \multicolumn{1}{c|}{1.00} & \multicolumn{1}{c|}{1.00} & \multicolumn{1}{c|}{0.99} & 1.00 & Free \\ \cline{3-11} 
\multicolumn{1}{c|}{} & \multicolumn{1}{c|}{} & \begin{tabular}[c]{@{}c@{}}$z$ \\ Redshift\end{tabular} & \multicolumn{7}{c|}{0} & Frozen \\ \cline{3-11} 
\multicolumn{1}{c|}{} & \multicolumn{1}{c|}{} & \begin{tabular}[c]{@{}c@{}}$kT$(keV)\\ Temperature\end{tabular} & \multicolumn{1}{c|}{0.67} & \multicolumn{1}{c|}{0.76} & \multicolumn{1}{c|}{0.70} & \multicolumn{1}{c|}{0.76} & \multicolumn{1}{c|}{0.71} & \multicolumn{1}{c|}{0.66} & 0.86 & Free (0.5, 1) \\ \cline{3-11} 
\multicolumn{1}{c|}{} & \multicolumn{1}{c|}{} & Normalization & \multicolumn{1}{c|}{0.092} & \multicolumn{1}{c|}{0.065} & \multicolumn{1}{c|}{0.075} & \multicolumn{1}{c|}{0.064} & \multicolumn{1}{c|}{0.047} & \multicolumn{1}{c|}{0.061} & 0.041 & Free \\ \cline{2-11} 
\multicolumn{1}{c|}{} & \multicolumn{1}{c|}{\multirow{2}{*}{power law}} & \begin{tabular}[c]{@{}c@{}}$\alpha$\\ Index\end{tabular} & \multicolumn{1}{c|}{1.51} & \multicolumn{1}{c|}{1.44} & \multicolumn{1}{c|}{1.50} & \multicolumn{1}{c|}{1.46} & \multicolumn{1}{c|}{1.58} & \multicolumn{1}{c|}{1.47} & 1.50 & Free (-3, 10) \\ \cline{3-11} 
\multicolumn{1}{c|}{} & \multicolumn{1}{c|}{} & Normalization & \multicolumn{1}{c|}{0.37} & \multicolumn{1}{c|}{0.34} & \multicolumn{1}{c|}{0.37} & \multicolumn{1}{c|}{0.35} & \multicolumn{1}{c|}{0.34} & \multicolumn{1}{c|}{0.29} & 0.36 & Free \\ \cline{2-11} 
\multicolumn{1}{c|}{} & \multicolumn{1}{c|}{tbabs} & \begin{tabular}[c]{@{}c@{}}$nH\rm\,(\times10^{20}cm^{-2})$\\ Column density\end{tabular} & \multicolumn{1}{c|}{4.4} & \multicolumn{1}{c|}{4.4} & \multicolumn{1}{c|}{4.4} & \multicolumn{1}{c|}{4.4} & \multicolumn{1}{c|}{4.4} & \multicolumn{1}{c|}{4.4} & 4.4 & Free \\ \hline
\multicolumn{1}{c|}{\multirow{42}{*}{Instrument}} & \multicolumn{1}{c|}{\multirow{29}{*}{gaussian}} & \begin{tabular}[c]{@{}c@{}}$E_l\rm\,(keV)$\\ Line energy\end{tabular} & \multicolumn{1}{c|}{9.80} & \multicolumn{1}{c|}{9.80} & \multicolumn{1}{c|}{9.57} & \multicolumn{1}{c|}{9.80} & \multicolumn{1}{c|}{9.80} & \multicolumn{1}{c|}{9.80} & 9.80 & Free (9.45, 9.8) \\ \cline{3-11} 
\multicolumn{1}{c|}{} & \multicolumn{1}{c|}{} & Normalization & \multicolumn{1}{c|}{1.78e-18} & \multicolumn{1}{c|}{1.78e-18} & \multicolumn{1}{c|}{0.00} & \multicolumn{1}{c|}{1.59e-18} & \multicolumn{1}{c|}{5.05e-18} & \multicolumn{1}{c|}{7.89e-19} & 2.34e-18 & Free \\ \cline{3-11} 
\multicolumn{1}{c|}{} & \multicolumn{1}{c|}{} & \begin{tabular}[c]{@{}c@{}}$E_l\rm\,(keV)$\\ Line energy\end{tabular} & \multicolumn{1}{c|}{8.85} & \multicolumn{1}{c|}{8.85} & \multicolumn{1}{c|}{8.85} & \multicolumn{1}{c|}{8.85} & \multicolumn{1}{c|}{9.10} & \multicolumn{1}{c|}{8.85} & 8.85 & Free (8.8, 9.15) \\ \cline{3-11} 
\multicolumn{1}{c|}{} & \multicolumn{1}{c|}{} & Normalization & \multicolumn{1}{c|}{4.19e-19} & \multicolumn{1}{c|}{4.34e-07} & \multicolumn{1}{c|}{7.20e-19} & \multicolumn{1}{c|}{5.17e-04} & \multicolumn{1}{c|}{7.89e-18} & \multicolumn{1}{c|}{8.63e-19} & 9.08e-04 & Free \\ \cline{3-11} 
\multicolumn{1}{c|}{} & \multicolumn{1}{c|}{} & \begin{tabular}[c]{@{}c@{}}$E_l\rm\,(keV)$\\ Line energy\end{tabular} & \multicolumn{1}{c|}{8.30} & \multicolumn{1}{c|}{8.30} & \multicolumn{1}{c|}{8.23} & \multicolumn{1}{c|}{8.30} & \multicolumn{1}{c|}{8.30} & \multicolumn{1}{c|}{8.20} & 8.23 & Free (8.15, 8.35) \\ \cline{3-11} 
\multicolumn{1}{c|}{} & \multicolumn{1}{c|}{} & Normalization & \multicolumn{1}{c|}{1.90e-05} & \multicolumn{1}{c|}{2.99e-18} & \multicolumn{1}{c|}{8.27e-04} & \multicolumn{1}{c|}{4.96e-19} & \multicolumn{1}{c|}{1.75e-04} & \multicolumn{1}{c|}{1.55e-18} & 1.39e-03 & Free \\ \cline{3-11} 
\multicolumn{1}{c|}{} & \multicolumn{1}{c|}{} & \begin{tabular}[c]{@{}c@{}}$E_l\rm\,(keV)$\\ Line energy\end{tabular} & \multicolumn{1}{c|}{8.60} & \multicolumn{1}{c|}{8.63} & \multicolumn{1}{c|}{8.62} & \multicolumn{1}{c|}{8.62} & \multicolumn{1}{c|}{8.66} & \multicolumn{1}{c|}{8.59} & 8.63 & Free (7.45, 8.75) \\ \cline{3-11} 
\multicolumn{1}{c|}{} & \multicolumn{1}{c|}{} & Normalization & \multicolumn{1}{c|}{5.26e-03} & \multicolumn{1}{c|}{5.46e-03} & \multicolumn{1}{c|}{4.22e-03} & \multicolumn{1}{c|}{4.35e-03} & \multicolumn{1}{c|}{4.57e-03} & \multicolumn{1}{c|}{5.48e-03} & 6.99e-03 & Free \\ \cline{3-11} 
\multicolumn{1}{c|}{} & \multicolumn{1}{c|}{} & \begin{tabular}[c]{@{}c@{}}$E_l\rm\,(keV)$\\ Line energy\end{tabular} & \multicolumn{1}{c|}{8.04} & \multicolumn{1}{c|}{8.09} & \multicolumn{1}{c|}{8.07} & \multicolumn{1}{c|}{8.10} & \multicolumn{1}{c|}{8.10} & \multicolumn{1}{c|}{8.07} & 8.08 & Free (7.8, 8.15) \\ \cline{3-11} 
\multicolumn{1}{c|}{} & \multicolumn{1}{c|}{} & Normalization & \multicolumn{1}{c|}{2.40e-03} & \multicolumn{1}{c|}{1.61e-03} & \multicolumn{1}{c|}{2.51e-03} & \multicolumn{1}{c|}{2.76e-03} & \multicolumn{1}{c|}{1.95e-03} & \multicolumn{1}{c|}{2.76e-03} & 1.57e-03 & Free \\ \cline{3-11} 
\multicolumn{1}{c|}{} & \multicolumn{1}{c|}{} & \begin{tabular}[c]{@{}c@{}}$E_l\rm\,(keV)$\\ Line energy\end{tabular} & \multicolumn{1}{c|}{7.47} & \multicolumn{1}{c|}{7.49} & \multicolumn{1}{c|}{7.49} & \multicolumn{1}{c|}{7.47} & \multicolumn{1}{c|}{7.50} & \multicolumn{1}{c|}{7.47} & 7.48 & Free (7.2, 7.6) \\ \cline{3-11} 
\multicolumn{1}{c|}{} & \multicolumn{1}{c|}{} & Normalization & \multicolumn{1}{c|}{3.45e-03} & \multicolumn{1}{c|}{3.51e-03} & \multicolumn{1}{c|}{4.91e-03} & \multicolumn{1}{c|}{4.33e-03} & \multicolumn{1}{c|}{4.48e-03} & \multicolumn{1}{c|}{4.54e-03} & 4.92e-03 & Free \\ \cline{3-11} 
\multicolumn{1}{c|}{} & \multicolumn{1}{c|}{} & \begin{tabular}[c]{@{}c@{}}$E_l\rm\,(keV)$\\ Line energy\end{tabular} & \multicolumn{1}{c|}{7.01} & \multicolumn{1}{c|}{6.97} & \multicolumn{1}{c|}{7.05} & \multicolumn{1}{c|}{7.02} & \multicolumn{1}{c|}{7.05} & \multicolumn{1}{c|}{7.04} & 6.95 & Free (6.9, 7.2) \\ \cline{3-11} 
\multicolumn{1}{c|}{} & \multicolumn{1}{c|}{} & Normalization & \multicolumn{1}{c|}{2.75e-03} & \multicolumn{1}{c|}{2.69e-03} & \multicolumn{1}{c|}{2.11e-03} & \multicolumn{1}{c|}{1.26e-03} & \multicolumn{1}{c|}{1.26e-03} & \multicolumn{1}{c|}{1.70e-03} & 8.12e-04 & Free \\ \cline{3-11} 
\multicolumn{1}{c|}{} & \multicolumn{1}{c|}{} & \begin{tabular}[c]{@{}c@{}}$E_l\rm\,(keV)$\\ Line energy\end{tabular} & \multicolumn{1}{c|}{6.86} & \multicolumn{1}{c|}{7.05} & \multicolumn{1}{c|}{6.91} & \multicolumn{1}{c|}{6.94} & \multicolumn{1}{c|}{6.97} & \multicolumn{1}{c|}{6.94} & 7.01 & Free (6.8, 7.1) \\ \cline{3-11} 
\multicolumn{1}{c|}{} & \multicolumn{1}{c|}{} & Normalization & \multicolumn{1}{c|}{9.23e-04} & \multicolumn{1}{c|}{1.37e-18} & \multicolumn{1}{c|}{9.54e-04} & \multicolumn{1}{c|}{6.49e-04} & \multicolumn{1}{c|}{1.20e-03} & \multicolumn{1}{c|}{1.12e-03} & 1.36e-03 & Free \\ \cline{3-11} 
\multicolumn{1}{c|}{} & \multicolumn{1}{c|}{} & \begin{tabular}[c]{@{}c@{}}$E_l\rm\,(keV)$\\ Line energy\end{tabular} & \multicolumn{1}{c|}{6.39} & \multicolumn{1}{c|}{6.41} & \multicolumn{1}{c|}{6.41} & \multicolumn{1}{c|}{6.41} & \multicolumn{1}{c|}{6.42} & \multicolumn{1}{c|}{6.40} & 6.41 & Free (6.2, 6.6) \\ \cline{3-11} 
\multicolumn{1}{c|}{} & \multicolumn{1}{c|}{} & Normalization & \multicolumn{1}{c|}{0.013} & \multicolumn{1}{c|}{0.011} & \multicolumn{1}{c|}{0.012} & \multicolumn{1}{c|}{0.012} & \multicolumn{1}{c|}{0.012} & \multicolumn{1}{c|}{0.011} & 0.012 & Free \\ \cline{3-11} 
\multicolumn{1}{c|}{} & \multicolumn{1}{c|}{} & \begin{tabular}[c]{@{}c@{}}$E_l\rm\,(keV)$\\ Line energy\end{tabular} & \multicolumn{1}{c|}{5.90} & \multicolumn{1}{c|}{5.94} & \multicolumn{1}{c|}{5.93} & \multicolumn{1}{c|}{5.92} & \multicolumn{1}{c|}{5.95} & \multicolumn{1}{c|}{5.90} & 5.92 & Free (5.7, 6) \\ \cline{3-11} 
\multicolumn{1}{c|}{} & \multicolumn{1}{c|}{} & Normalization & \multicolumn{1}{c|}{1.76e-03} & \multicolumn{1}{c|}{1.76e-03} & \multicolumn{1}{c|}{1.46e-03} & \multicolumn{1}{c|}{7.50e-04} & \multicolumn{1}{c|}{9.91e-04} & \multicolumn{1}{c|}{3.12e-04} & 1.64e-03 & Free \\ \cline{3-11} 
\multicolumn{1}{c|}{} & \multicolumn{1}{c|}{} & \begin{tabular}[c]{@{}c@{}}$E_l\rm\,(keV)$\\ Line energy\end{tabular} & \multicolumn{1}{c|}{5.44} & \multicolumn{1}{c|}{5.55} & \multicolumn{1}{c|}{5.40} & \multicolumn{1}{c|}{5.42} & \multicolumn{1}{c|}{5.55} & \multicolumn{1}{c|}{5.55} & 5.47 & Free (5.3, 5.6) \\ \cline{3-11} 
\multicolumn{1}{c|}{} & \multicolumn{1}{c|}{} & Normalization & \multicolumn{1}{c|}{8.67e-04} & \multicolumn{1}{c|}{5.10e-19} & \multicolumn{1}{c|}{2.67e-04} & \multicolumn{1}{c|}{6.34e-04} & \multicolumn{1}{c|}{1.25e-17} & \multicolumn{1}{c|}{2.70e-19} & 8.99e-04 & Free \\ \cline{3-11} 
\multicolumn{1}{c|}{} & \multicolumn{1}{c|}{} & \begin{tabular}[c]{@{}c@{}}$E_l\rm\,(keV)$\\ Line energy\end{tabular} & \multicolumn{1}{c|}{4.50} & \multicolumn{1}{c|}{4.50} & \multicolumn{1}{c|}{4.52} & \multicolumn{1}{c|}{4.52} & \multicolumn{1}{c|}{4.52} & \multicolumn{1}{c|}{4.52} & 4.52 & Free (4.49, 4.53) \\ \cline{3-11} 
\multicolumn{1}{c|}{} & \multicolumn{1}{c|}{} & Normalization & \multicolumn{1}{c|}{3.26e-04} & \multicolumn{1}{c|}{8.63e-04} & \multicolumn{1}{c|}{1.61e-18} & \multicolumn{1}{c|}{1.13e-04} & \multicolumn{1}{c|}{6.49e-04} & \multicolumn{1}{c|}{2.29e-18} & 8.60e-04 & Free \\ \cline{3-11} 
\multicolumn{1}{c|}{} & \multicolumn{1}{c|}{} & \begin{tabular}[c]{@{}c@{}}$E_l\rm\,(keV)$\\ Line energy\end{tabular} & \multicolumn{1}{c|}{3.67} & \multicolumn{1}{c|}{3.68} & \multicolumn{1}{c|}{3.68} & \multicolumn{1}{c|}{3.66} & \multicolumn{1}{c|}{3.68} & \multicolumn{1}{c|}{3.67} & 3.72 & Free (3.6, 3.8) \\ \cline{3-11} 
\multicolumn{1}{c|}{} & \multicolumn{1}{c|}{} & Normalization & \multicolumn{1}{c|}{1.08e-03} & \multicolumn{1}{c|}{1.63e-03} & \multicolumn{1}{c|}{1.76e-03} & \multicolumn{1}{c|}{1.10e-03} & \multicolumn{1}{c|}{1.20e-03} & \multicolumn{1}{c|}{1.08e-03} & 1.07e-03 & Free \\ \cline{3-11} 
\multicolumn{1}{c|}{} & \multicolumn{1}{c|}{} & \begin{tabular}[c]{@{}c@{}}$E_l\rm\,(keV)$\\ line energy\end{tabular} & \multicolumn{1}{c|}{1.48} & \multicolumn{1}{c|}{1.49} & \multicolumn{1}{c|}{1.49} & \multicolumn{1}{c|}{1.49} & \multicolumn{1}{c|}{1.48} & \multicolumn{1}{c|}{1.48} & 1.49 & Free (1.47, 1.5) \\ \cline{3-11} 
\multicolumn{1}{c|}{} & \multicolumn{1}{c|}{} & Normalization & \multicolumn{1}{c|}{4.94e-03} & \multicolumn{1}{c|}{5.23e-03} & \multicolumn{1}{c|}{4.66e-03} & \multicolumn{1}{c|}{5.48e-03} & \multicolumn{1}{c|}{3.19e-03} & \multicolumn{1}{c|}{6.22e-03} & 2.99e-03 & Free \\ \cline{3-11} 
\multicolumn{1}{c|}{} & \multicolumn{1}{c|}{} & \begin{tabular}[c]{@{}c@{}}$\sigma\rm\,(keV)$\\ Line width (for all lines)\end{tabular} & \multicolumn{7}{c|}{0.0001} & Frozen \\ \cline{2-11} 
\multicolumn{1}{c|}{} & \multicolumn{1}{c|}{\multirow{3}{*}{\begin{tabular}[c]{@{}c@{}}exponential\\ modification\end{tabular}}} & Amplitude & \multicolumn{1}{c|}{-1.67e-10} & \multicolumn{1}{c|}{-1.40e-10} & \multicolumn{1}{c|}{-9.69e-09} & \multicolumn{1}{c|}{-1.02e-07} & \multicolumn{1}{c|}{-7.45e-11} & \multicolumn{1}{c|}{2.46e-10} & -1.27e-08 & Free (-2, 2) \\ \cline{3-11} 
\multicolumn{1}{c|}{} & \multicolumn{1}{c|}{} & \begin{tabular}[c]{@{}c@{}}$f$\\ Factor\end{tabular} & \multicolumn{1}{c|}{-2.10} & \multicolumn{1}{c|}{-2.12} & \multicolumn{1}{c|}{-1.68} & \multicolumn{1}{c|}{-1.49} & \multicolumn{1}{c|}{-2.25} & \multicolumn{1}{c|}{-2.03} & -1.75 & Free (-4, 4) \\ \cline{3-11} 
\multicolumn{1}{c|}{} & \multicolumn{1}{c|}{} & \begin{tabular}[c]{@{}c@{}}$E_c$$\rm(keV)$\\ Start energy\end{tabular} & \multicolumn{1}{c|}{0.05} & \multicolumn{1}{c|}{0.05} & \multicolumn{1}{c|}{0.05} & \multicolumn{1}{c|}{0.05} & \multicolumn{1}{c|}{0.05} & \multicolumn{1}{c|}{0.05} & 0.05 & Free \\ \cline{2-11} 
\multicolumn{1}{c|}{} & \multicolumn{1}{c|}{\multirow{6}{*}{\begin{tabular}[c]{@{}c@{}}broken power law,\\ 2 break energies\end{tabular}}} & $\Gamma_1$ & \multicolumn{1}{c|}{0.06} & \multicolumn{1}{c|}{0.07} & \multicolumn{1}{c|}{0.09} & \multicolumn{1}{c|}{0.11} & \multicolumn{1}{c|}{0.18} & \multicolumn{1}{c|}{0.07} & 0.14 & Free (-3, 10) \\ \cline{3-11} 
\multicolumn{1}{c|}{} & \multicolumn{1}{c|}{} & \begin{tabular}[c]{@{}c@{}}$E_{break,1}$\\ $\rm(keV)$\end{tabular} & \multicolumn{1}{c|}{3.96} & \multicolumn{1}{c|}{5.26} & \multicolumn{1}{c|}{4.71} & \multicolumn{1}{c|}{8.89} & \multicolumn{1}{c|}{8.92} & \multicolumn{1}{c|}{6.23} & 96.90 & Free \\ \cline{3-11} 
\multicolumn{1}{c|}{} & \multicolumn{1}{c|}{} & $\Gamma_2$ & \multicolumn{1}{c|}{0.14} & \multicolumn{1}{c|}{0.19} & \multicolumn{1}{c|}{0.17} & \multicolumn{1}{c|}{-2.50} & \multicolumn{1}{c|}{-2.50} & \multicolumn{1}{c|}{0.28} & 0.03 & Free (-3, 10) \\ \cline{3-11} 
\multicolumn{1}{c|}{} & \multicolumn{1}{c|}{} & \begin{tabular}[c]{@{}c@{}}$E_{break,2}$\\ $\rm(keV)$\end{tabular} & \multicolumn{1}{c|}{8.89} & \multicolumn{1}{c|}{9.35} & \multicolumn{1}{c|}{18.08} & \multicolumn{1}{c|}{12.79} & \multicolumn{1}{c|}{10.84} & \multicolumn{1}{c|}{8.87} & 9.50 & Free \\ \cline{3-11} 
\multicolumn{1}{c|}{} & \multicolumn{1}{c|}{} & $\Gamma_3$ & \multicolumn{1}{c|}{7.71} & \multicolumn{1}{c|}{-2.31} & \multicolumn{1}{c|}{9.15} & \multicolumn{1}{c|}{2.00} & \multicolumn{1}{c|}{-1.54} & \multicolumn{1}{c|}{9.50} & 0.23 & Free (-3, 10) \\ \cline{3-11} 
\multicolumn{1}{c|}{} & \multicolumn{1}{c|}{} & Normalization & \multicolumn{1}{c|}{0.20} & \multicolumn{1}{c|}{0.19} & \multicolumn{1}{c|}{0.20} & \multicolumn{1}{c|}{0.21} & \multicolumn{1}{c|}{0.27} & \multicolumn{1}{c|}{0.20} & 0.24 & Free \\ \cline{2-11} 
\multicolumn{1}{c|}{} & \multicolumn{1}{c|}{\multirow{2}{*}{power law}} & \begin{tabular}[c]{@{}c@{}}$\alpha$\\ Index\end{tabular} & \multicolumn{1}{c|}{1.24} & \multicolumn{1}{c|}{2.05} & \multicolumn{1}{c|}{0.80} & \multicolumn{1}{c|}{2.34} & \multicolumn{1}{c|}{5.85} & \multicolumn{1}{c|}{1.86} & 1.73 & Free (-3, 10) \\ \cline{3-11} 
\multicolumn{1}{c|}{} & \multicolumn{1}{c|}{} & Normalization & \multicolumn{1}{c|}{3.77e-03} & \multicolumn{1}{c|}{0.05} & \multicolumn{1}{c|}{2.65e-03} & \multicolumn{1}{c|}{0.01} & \multicolumn{1}{c|}{0.02} & \multicolumn{1}{c|}{0.05} & 0.06 & Free \\ \cline{2-11} 
\multicolumn{1}{c|}{} & \multicolumn{1}{c|}{\multirow{2}{*}{power law}} & \begin{tabular}[c]{@{}c@{}}$\alpha$\\ Index\end{tabular} & \multicolumn{1}{c|}{2.78} & \multicolumn{1}{c|}{6.00} & \multicolumn{1}{c|}{2.35} & \multicolumn{1}{c|}{8.83} & \multicolumn{1}{c|}{6.00} & \multicolumn{1}{c|}{1.21} & 6.00 & \begin{tabular}[c]{@{}c@{}}Free (-3, 10)\\ except TM5 \& TM7\end{tabular} \\ \cline{3-11} 
\multicolumn{1}{c|}{} & \multicolumn{1}{c|}{} & \begin{tabular}[c]{@{}c@{}}$K$\\ Normalization\end{tabular} & \multicolumn{1}{c|}{0.03} & \multicolumn{1}{c|}{0.00} & \multicolumn{1}{c|}{0.02} & \multicolumn{1}{c|}{9.68e-04} & \multicolumn{1}{c|}{0.00} & \multicolumn{1}{c|}{3.22e-05} & 0.00 & \begin{tabular}[c]{@{}c@{}}Free (-3, 10) \\ except TM5 \& TM7\end{tabular} \\ \hline
\end{tabular}%
\label{all+model_param_table}
}
 \caption{All best fit parameter values in eFEDS blank sky models. Unless with a specified range, ``Free" means we allow the range of the floating parameters are allowed to be any non negative number. The instrument parameters floating ranges follow eROSITA released FWC model. All parameter are in units acceptable by \texttt{xspec}.}
\end{table*}

\bigskip

\section{\label{sec:plawind}The Impact of Using Different CXB Power Law Indices on DM Constraint}

\par
\begin{figure*}[hbt]
    \includegraphics[width=0.6\linewidth]{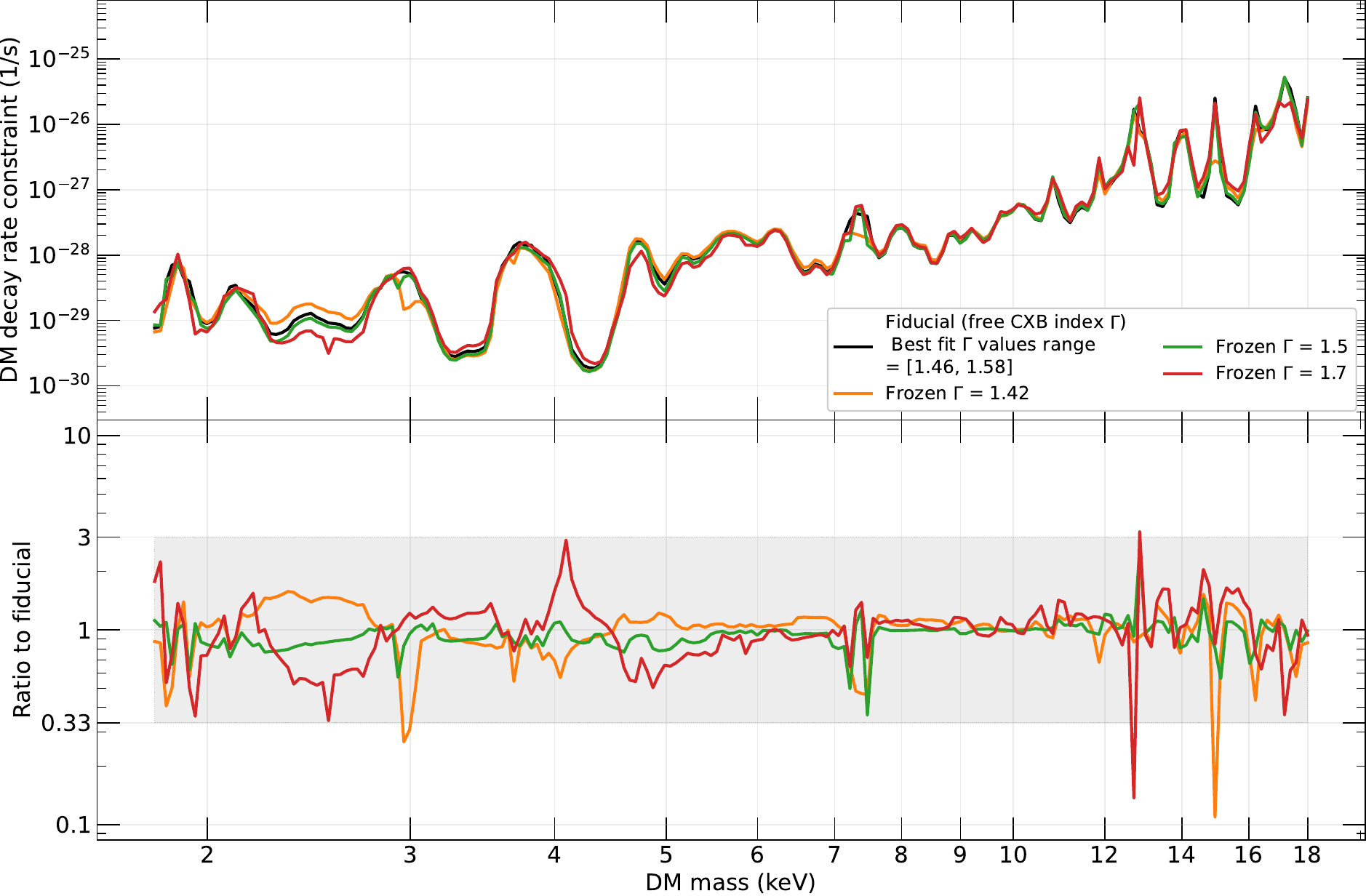}
    \caption{Comparison of DM constraint results using CXB model with different power law indices. Top: DM decay rate constraint results from using CXB power law indices fixed to 3 different values, compared to our fiducial result where the power law index is allowed to take any value. Bottom: ratios of DM decay rate constraint results to our fiducial result. Gray color in the bottom panel marks where the ratio is within factor of 3. }
    \label{fig:3plaw}
\end{figure*}

 We investigate how different models of CXB could impact our DM constraint result. Since CXB is modeled by a power law and we allow all normalization to float freely in our DM search procedure, the only choice in CXB parameter is the index of power law. In figure~\ref{fig:3plaw} we show 4 constraint results with different choices of power law indices: (1) our fiducial result, where power law index is unfrozen (free) in initial spectral fitting and DM search, with the best values of the 7 TMs falling between 1.46-1.58; (2) power law index frozen at 1.42, which is reported in XMM-Newton papers and suitable for hard X-ray sky~\cite{Bulbul:2011ku}; (3) frozen at 1.5 which is close to the average of our best fit null models for 7 individual TMs; (4) frozen at 1.7 which is reported in eROSITA diffuse sky study and suitable for soft X-ray sky~\cite{Ponti:2022nix}. As shown in the bottom panel of figure~\ref{fig:3plaw}, we find that the difference choices of CXB power law model indices have minor impact on our DM constraint result. The ratios of DM constraints from different CXB models to our fiducial result are within factor of 3, except in narrow ranges of energy where the fit is problematic for all models.

\end{document}